\documentclass[english, prb, reprint, superscriptaddress,aps]{revtex4-2}
\usepackage[utf8]{inputenc}    
\usepackage[T1]{fontenc} 
\usepackage{graphicx}
\usepackage{microtype} 
\usepackage{amsmath}  
\usepackage{amssymb} 
\usepackage{xcolor} 
\usepackage{ulem}
\usepackage{times}
\usepackage[colorlinks=true,citecolor=blue,linkcolor=blue]{hyperref}
\definecolor{RED}{rgb}{1,0,0}

\usepackage[english]{babel}
\DeclareUnicodeCharacter{00E9}{\'{e}}    
\DeclareUnicodeCharacter{00E1}{\'{a}}   
\DeclareUnicodeCharacter{00F3}{\'{o}}
\makeatletter

\renewcommand{\fnum@figure}{Fig.~ \thefigure}
\makeatother

\begin{document}

\title{Symmetry enforced quantum spin Hall effect in Altermagnets}

\author{Fanzheng Chen}
\affiliation{College of Sciences, Northeastern University, Shenyang 110819, China}

\author{Lixin Zhang}
\affiliation{School of Physics and Optoelectronics, Shandong Normal University, Jinan, 250358, China}

\author{Shuaishuai Niu}
\affiliation{College of Sciences, Northeastern University, Shenyang 110819, China}

\author{Junfeng Ren}
\affiliation{School of Physics and Optoelectronics, Shandong Normal University, Jinan, 250358, China}

\author{Weijiang Gong}
\affiliation{College of Sciences, Northeastern University, Shenyang 110819, China}

\author{Xiangru Kong}
\email{Contact author: kongxiangru@neu.edu.cn}
\affiliation{College of Sciences, Northeastern University, Shenyang 110819, China}

\begin{abstract}
The quantum spin Hall effect (QSHE) has attracted widespread attention due to its dissipationless transport, which is protected by non-trivial topological invariants and helical edge states. Because even weak magnetic disorder can destroy the stability of topological quantum states, current research on the QSHE has primarily focused on non-magnetic materials. In this work, we extend the research scope of the QSHE to altermagnets. We establish the relevant symmetry constraints and identify all magnetic point groups that can realize the altermagnetic QSHE. Symmetry analysis reveals that pronounced spin-valley locking or spin-valley-layer locking universally exists in these systems. The concerted interaction between band inversion and spin-valley locking collectively gives rise to the helical edge states. Using first-principles calculations and theoretical models, we demonstrate that monolayer $\text{Nb}_2\text{SeTeO}$ exhibits an altermagnetic QSHE characterized by spin-valley locking, while bilayer $\text{Hf}_3\text{Se}_3\text{Te}_2$ manifests an altermagnetic QSHE featuring spin-valley-layer locking. This work clarifies the intrinsic symmetry correlation between altermagnetism and quantum spin Hall topological phases, providing a brand-new theoretical perspective and research platform for exploring magnetic topological systems and developing next-generation spintronic devices.
\end{abstract}
\maketitle

\begin{figure*}[htbp]
\centering
\includegraphics[width=0.95\linewidth]{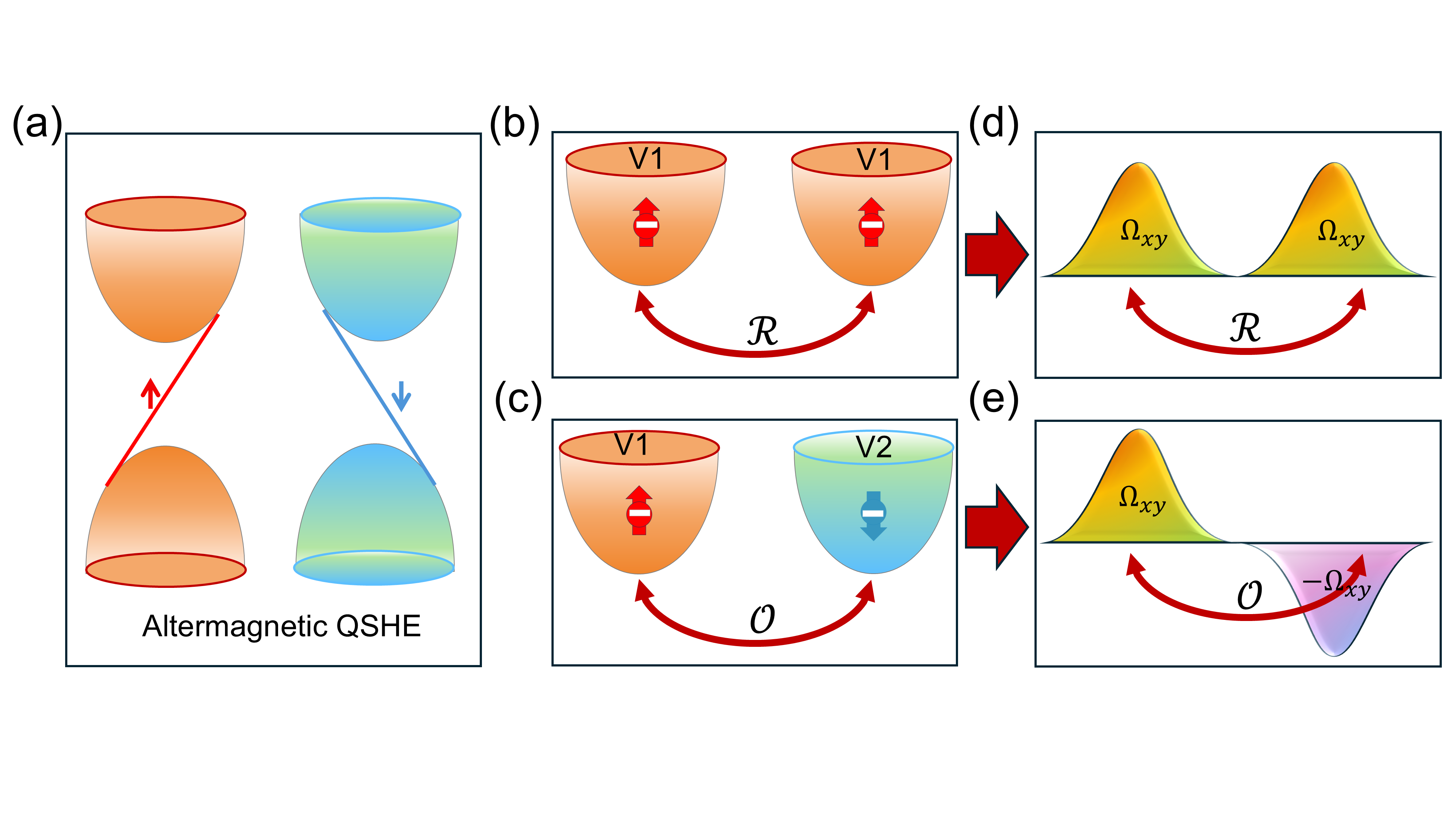} 
\vspace{-5em}
\caption{Schematic of QSHE with SVL.
(a) Helical edge states of altermagnetic QSHE.
(b) Altermagnets with two valleys $V_1$ and $V_2$ protected by (magnetic) crystalline symmetry $\mathcal{R}$.
(c) The two valleys $V_1$ and $V_2$ are linked by (magnetic) crystalline symmetry $\mathcal{O}$.
(d) Symmetry $\mathcal{R}$ guarantees nonvanishing Berry curvatures for $V_1$ and $V_2$.
(e) Symmetry $\mathcal{O}$ enforces the total Berry curvature to vanish strictly.}
\label{fig:label1}
\end{figure*}

\textit{Introduction}---The quantum spin Hall effect (QSHE) represents a class of typical topological phases of matter~\cite{Bansil2016,Hasan2010,Qi2011,Breunig2022,Xiao2021}, which has attracted widespread attention in the fields of condensed matter physics and materials science due to its non-trivial topological invariants and helical edge states protected by time-reversal symmetry (TRS)~\cite{Kane2005,Kane2005graphene,Maciejko2009}. The edge conduction channels of this topological quantum state exhibit backscattering-suppressed characteristics~\cite{Kimme2016,Jack2020}, providing an important physical platform for the construction of low-power, high-speed next-generation spintronics devices. The conventional QSHE is mainly realized in non-magnetic topological insulators (TI), such as $\text{HgTe/CdTe}$ quantum wells and monolayer $\text{WTe}_2$~\cite{Bernevig2006,Konig2007,Tang2017}. Since QSHE is protected by TRS, even weak magnetic disorder can destroy the stability of the topological quantum state~\cite{Xu2006,Hattori2011J}. Therefore, how to extend QSHE to magnetic systems and achieve the cooperative modulation of topological phases relying on magnetic order is an important frontier scientific issue in condensed matter physics~\cite{Niu2020,Jiang20251,zou2024}.

In recent years, altermagnetism has been proposed as the third fundamental magnetic state that combines the advantages of spin splitting in ferromagnets (FMs) and zero net magnetization in antiferromagnets (AFMs)~\cite{Smejkal2022_PRX,Smejkal2022_PRX2,Mazin2022,Mazin2023,GonzalezHernandez2021}. This unique property is protected by crystal symmetry~\cite{smejkal2022B}. Due to this characteristic, altermagnetism gives rise to numerous ferromagnet-like physical effects, such as the anomalous Hall effect, piezomagnetic effect, giant magnetoresistance, and tunneling magnetoresistance~\cite{Sinova2020,Dev2025,Ma2021,Tomas2022}. Recent studies have shown that altermagnets (AMs) can host QSHE carrying helical edge states~\cite{Venderbos2025,Niu2025l,Jiang2025,Yao2026}.Unlike conventional non-magnetic QSHE, the topological invariant of altermagnetic QSHE is described by $C_s$~\cite{Chen2026_arXiv2,Bernardo2025,Feng2025_arXiv}, and this topological quantum state remains stable under variations of interlayer stacking~\cite{TAN20262196,Chen2026}, exhibiting robust characteristics. Furthermore, while conventional QSHE supports only a single pair of helical edge states, altermagnetic QSHE can break through this limitation to support multiple pairs of helical edge states~\cite{Liu2026}. Although AMs show great application potential in the field of topological phases~\cite{Zou2025,Zhang2024,Yang2025,Liu2025}, research on the intrinsic correlation between AMs and QSHE remains scarce. In addition, the number of materials currently proposed capable of hosting altermagnetic QSHE remains limited~\cite{Hu2026,Jing2026_CJP,Feng2024,Fu2025_arXiv,Ma2021_JPCM}. How to systematically elucidate the realization conditions of the QSHE in AMs from a symmetry perspective and identify real materials capable of hosting this effect is a critical theoretical challenge that urgently needs to be solved.

Here,  we establish the relevant symmetry constraints and identify all magnetic point groups (MPGs) that can realize the two dimensional (2D) altermagnetic QSHE. Symmetry analysis demonstrates that pronounced spin-valley locking (SVL) or spin-valley-layer locking (SVLL) universally exist in these systems. On this basis, we reveal the cooperative microscopic mechanism between band inversion and SVL, and we prove that this mechanism is the physical origin for the generation of helical edge states. To verify these theoretical predictions, we select monolayer $\text{Nb}_2\text{SeTeO}$ and bilayer $\text{Hf}_3\text{Se}_3\text{Te}_2$ as candidate materials, combining first-principles calculations and theoretical model analysis. The results show that the monolayer $\text{Nb}_2\text{SeTeO}$ exhibits altermagnetic spin splitting accompanied by a distinct SVL behavior. Under spin-orbit coupling (SOC), the system opens an energy gap at the Weyl points and induces a pair of topologically protected helical edge states, which are protected by crystal symmetry rather than TRS. In contrast, the monolayer $\text{Hf}_3\text{Se}_3\text{Te}_2$ is a typical Chern insulator with $C = 1$, while the bilayer $\text{Hf}_3\text{Se}_3\text{Te}_2$ exhibits spin splitting driven by SVLL, which can similarly open a gap at the Weyl points under SOC and form a pair of helical edge states. This work not only clarifies the intrinsic symmetry correlation between altermagnetic order and QSH topological phases, but also provides a solid theoretical basis and a brand-new research paradigm for exploring magnetic topological materials and designing low-power spintronic devices.

\begin{table}[htbp]
\centering
\caption{MPG capable of hosting altermagnetic QSHE. The groups allowing for SVLL are highlighted by red.The numbering of MPG follows Ref.~\cite{2022MSG}.}
\label{tab:mlpg}
\begin{tabular}{ll}
\hline
Lattice & MPG \\
\hline
Oblique & \textcolor{red}{3.1.6($2.1$)}, 4.1.9($m.1$), 5.1.12($2/m.1$) \\
Rectangular & \textcolor{red}{6.1.17($222.1$)}, 7.1.20($mm2.1$), 8.1.24($mmm.1$) \\
Square & 9.3.31($4'$), \textcolor{red}{10.3.34($\bar4'$)}, 11.3.37($4'/m$), \\
& 12.3.42($4'2'2$), 13.3.46($4'm'm$), 14.3.50($\bar4'2'm$), \\
& \textcolor{red}{14.4.51($\bar4'm'2$)}, 15.4.56($4'/mm'm$) \\
Hexagonal & \textcolor{red}{18.1.65($32.1$)}, 19.1.68($3m.1$), 20.1.71($\bar3m.1$),\\
& \textcolor{red}{24.1.87($622.1$)}, 25.1.91($6mm.1$),\\
& 26.1.95($\bar{6}m2.1$), 27.1.100($6/mmm.1$) \\
\hline
\end{tabular}
\end{table}

\textit{Altermagnetic QSHE and symmetry analysis}---The QSHE was originally proposed in graphene~\cite{Kane2005graphene}. Graphene hosts a pair of energy-degenerate valleys at the $K$ and $K'$ points of the Brillouin zone, where the intrinsic SOC induces a topological band inversion, giving rise to a pair of helical edge states protected by the topological invariant $Z_2$~\cite{Kane2005}.

Analogously, we consider a class of 2D AMs featuring a well-defined valley structure~\cite{Ma1,Ma2,Ma3,Ma4}. Under the influence of altermagnetic order, the TRS of the system is broken, causing valleys with opposite spins to appear in pairs and naturally establishing the SVL characteristic~\cite{mazin20231, guo2023, tan2025, shi2026, xu2025L}. Upon the introduction of SOC, a topological band inversion occurs within each valley channel~\cite{Zhang2009}. Due to spin-up and spin-down valley channels simultaneously exist in pairs with opposite spin polarizations, they contribute edge states with opposite chiralities, whose topological properties are characterized by the spin Chern number $C_s$. The overlap of two edge channels carrying opposite chirality yields the altermagnetic QSHE, as shown in Fig. 1(a).

Next, we deduce the symmetry constraints for the QSHE induced by SVL. To simplify the discussion, we assume that the system contains only two valleys, denoted as $V_1$ and $V_2$ [see Fig. 1(b) and Fig. 1(c)], and that the Néel vector is oriented perpendicular to the plane. in such case, the directions of the spin polarization of the valley electrons for the systems without SOC must be along the z direction; furthermore, the  SOC does not alter this spin polarization direction. Consequently, the symmetry of the altermagnetic system can be divided into two subsets: the set of operators $\mathcal{O}$ that keeps the valley index invariant, and the set of operators $\mathcal{R}$ that interchanges the two valleys, as follows:
\begin{equation}
\mathcal{R}V_{1(2)} = V_{1(2)},\qquad \mathcal{O}V_{1(2)} = V_{2(1)}.
\label{eq:1}
\end{equation}

To realize the QSHE, the Berry curvature of each valley must be non-zero, while the total Berry curvature must be strictly zero [see Fig. 1(d) and Fig. 1(e)]. Therefore, any symmetry operation in $\mathcal{R}$ and $\mathcal{O}$ should possess the following characteristics:
\begin{equation}
\mathcal{R}\Omega_{xy}\mathcal{R}^{-1} = \Omega_{xy},\qquad \mathcal{O}\Omega_{xy}\mathcal{O}^{-1} = -\Omega_{xy}.
\label{eq:2}
\end{equation}

Since both spin and Berry curvature are axial vectors, they exhibit identical properties under symmetry operations. Therefore, the spin satisfies the following relations:
\begin{equation}
\mathcal{R}\hat{s}_z\mathcal{R}^{-1} = \hat{s}_z,\qquad \mathcal{O}\hat{s}_z\mathcal{O}^{-1} = -\hat{s}_z.
\label{eq:3}
\end{equation}

The altermagnetic QSHE with SVLL can be realized~\cite{Liu2026L,Zhang2024}, 
provided that the two sublattices of the system belong to different atomic layers and all symmetry operations within $\mathcal{R}$ and $\mathcal{O}$ satisfy the symmetry relations:
\begin{equation}
    \mathcal{R}\mathcal{L}_{1(2)} = \mathcal{L}_{1(2)}, \quad
    \mathcal{O}\mathcal{L}_{1(2)} = \mathcal{L}_{2(1)}.
\label{eq:4}
\end{equation}

Where $\mathcal{L}$ denotes the layer index. This symmetry configuration perfectly matches the unique spin-splitting characteristics of AMs. The magnetic properties of materials are generally described by  magnetic groups or spin groups~\cite{Elcoro2021,Chen2024,Xiao2024,Jiang2024}. The key difference is that magnetic groups consider SOC. while spin groups explicitly ignore SOC. Given that altermagnetic QSHE can only be physically realized in the presence of SOC, the use of MPG to analyze the topological classification is both sufficient and efficient. To this end, we perform a comprehensive group-theoretic classification of MPG to identify those capable of hosting the altermagnetic QSHE~\cite{2022MSG}. Via comprehensive symmetry screening, we obtain 21 MPG that can realize the altermagnetic QSHE, among which six exhibit SVLL behavior. Full details of the MPG analysis and  associated magnetic layer groups are given in the supplementary material~\cite{SM_Note}. The results of this investigation are summarized in Table 1, providing a definitive guideline for the experimental search and synthesis of such candidate topological materials.

\begin{figure}[htbp]
\centering
\includegraphics[width=3.3in]{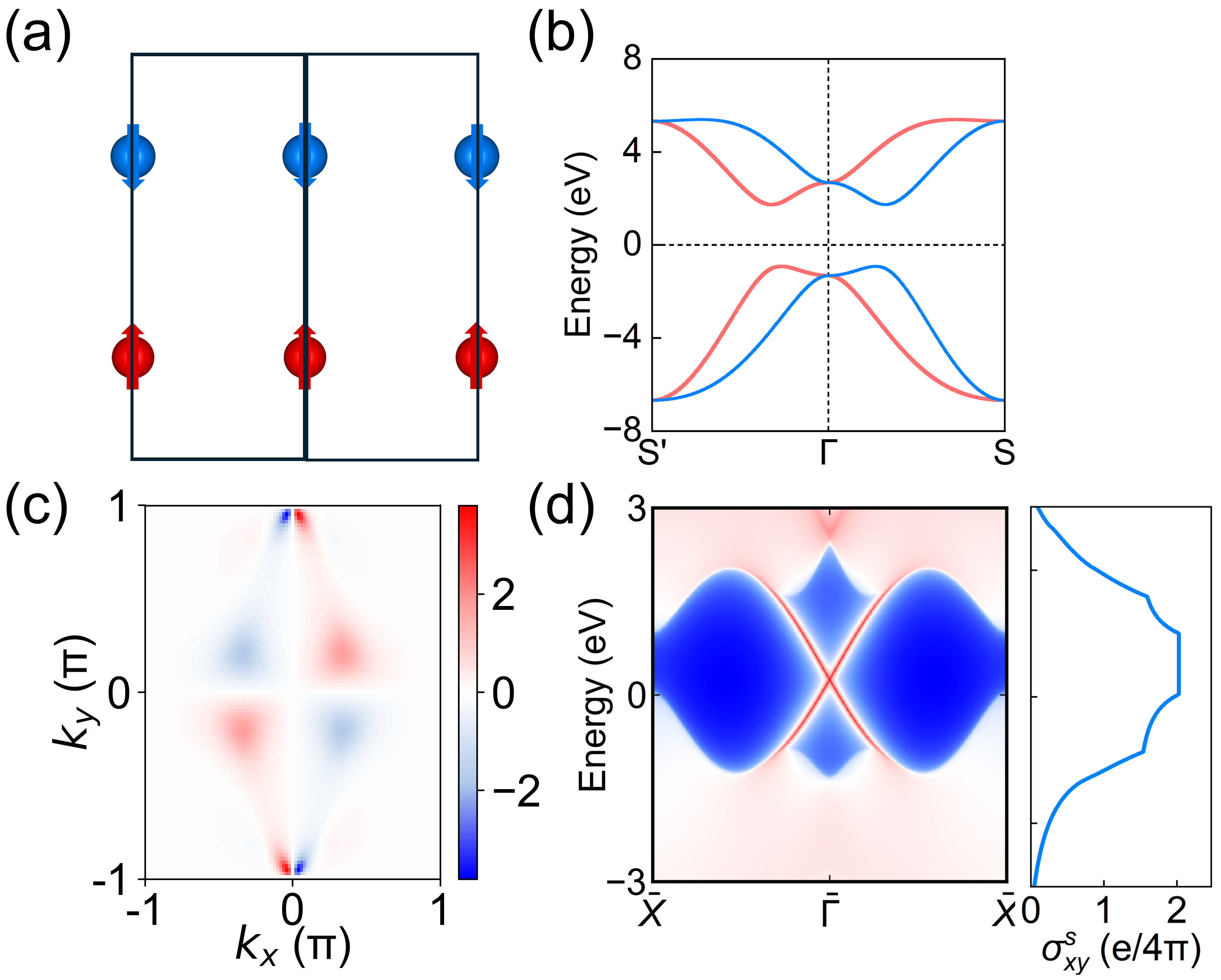} 
\vspace{-1em}%
\caption{(a) Side view of the altermagnetic QSHE lattice model.
(b) Band structure of the model, showing altermagnetic spin
splitting. The Brillouin zone is presented in Supplemental
Material.
(c) Berry curvature of the model. 
(d) Edge states and SHC of the model. Here, we set $t_1$=0.265, $t_2$=0.165, $t_3$=-0.2, $t_4$=-0.145, $r_1$=-0.29, $r_2$=0.42, $r_3$=-0.27, and $e_1$=$-e_2$= 0.5 in (b).}
\label{fig:label2}
\end{figure}

\textit{Lattice model}---We construct a minimal lattice model for altermagnetic QSHE systems within the MagneticTB Software~\cite{Zhang2022_CPC}. An rectangular lattice of  MPG 6.1.17($222.1$) is chosen (see Table I). Each unit cell hosts two lattice sites positioned at $(0,0,\tfrac14)$ and $(0,0,\tfrac34)$, where the two sites carry distinct spin orientations, as illustrated in Fig. 2(a). The lattice Hamiltonian respecting this symmetry can be formulated as:

\begin{equation}
\begin{aligned}
H &= \frac{\varepsilon_A+\varepsilon_B}{2}\tau_0\otimes\sigma_0+\frac{\varepsilon_A-\varepsilon_B}{2}\tau_z\otimes\sigma_z \\
  &\quad -8r_1\tau_x\otimes\sigma_z-8t_3\tau_y\otimes\sigma_z+8t_1\tau_x\otimes\sigma_0
\label{eq:5}
\end{aligned}
\end{equation}

With $\varepsilon_A = 4e_2 + 8r_2\cos k_y + 8t_2\cos k_x$ and $\varepsilon_B = 4e_1 + 8r_3\cos k_y + 8t_4\cos k_x$. Here, $\tau$ and $\sigma$ denote Pauli matrices acting on the site and spin spaces, respectively; $e_1=-e_2$ stands for the exchange term associated with AFM order.
The parameters $t_1$, $t_2$, $t_3$, and $t_4$ denote the nearest-neighbor (NN) hopping integrals, while $r_1$, $r_2$, $r_3$ correspond to the next-nearest-neighbor (NNN) hopping parameters. A nonzero $r_1$ is required to realize the altermagnetic spin splitting. The Hamiltonian calculations faithfully reproduce the characteristic band structure of SVL AMs [see Fig. 2(b)], and the features of the Berry curvature agree well with our symmetry analysis [see Fig. 2(c)]. The computed edge states exhibit helical edge states characteristic of the QSHE [see Fig. 2(d)]. The quantized SHC confirms the nontrivial topological nature of our lattice model with $C_\mathrm{s}=1$, where $C_\mathrm{s}=(C_\uparrow-C_\downarrow)/2$ and $C_\uparrow$, $C_\downarrow$ are obtained by integrating Berry curvatures of all occupied bands of each spin sector separately.

In addition to establishing the design principles, identifying concrete candidate materials is of equal importance. We demonstrate monolayer $\text{Nb}_2\text{SeTeO}$  and bilayer $\text{Hf}_3\text{Se}_3\text{Te}_2$ as candidate materials to validate our theoretical conclusions. Additional candidate materials can be found in the Supplementary Material~\cite{SM_Note}.

\begin{figure}[htbp]
\centering
\includegraphics[width=3.3in]{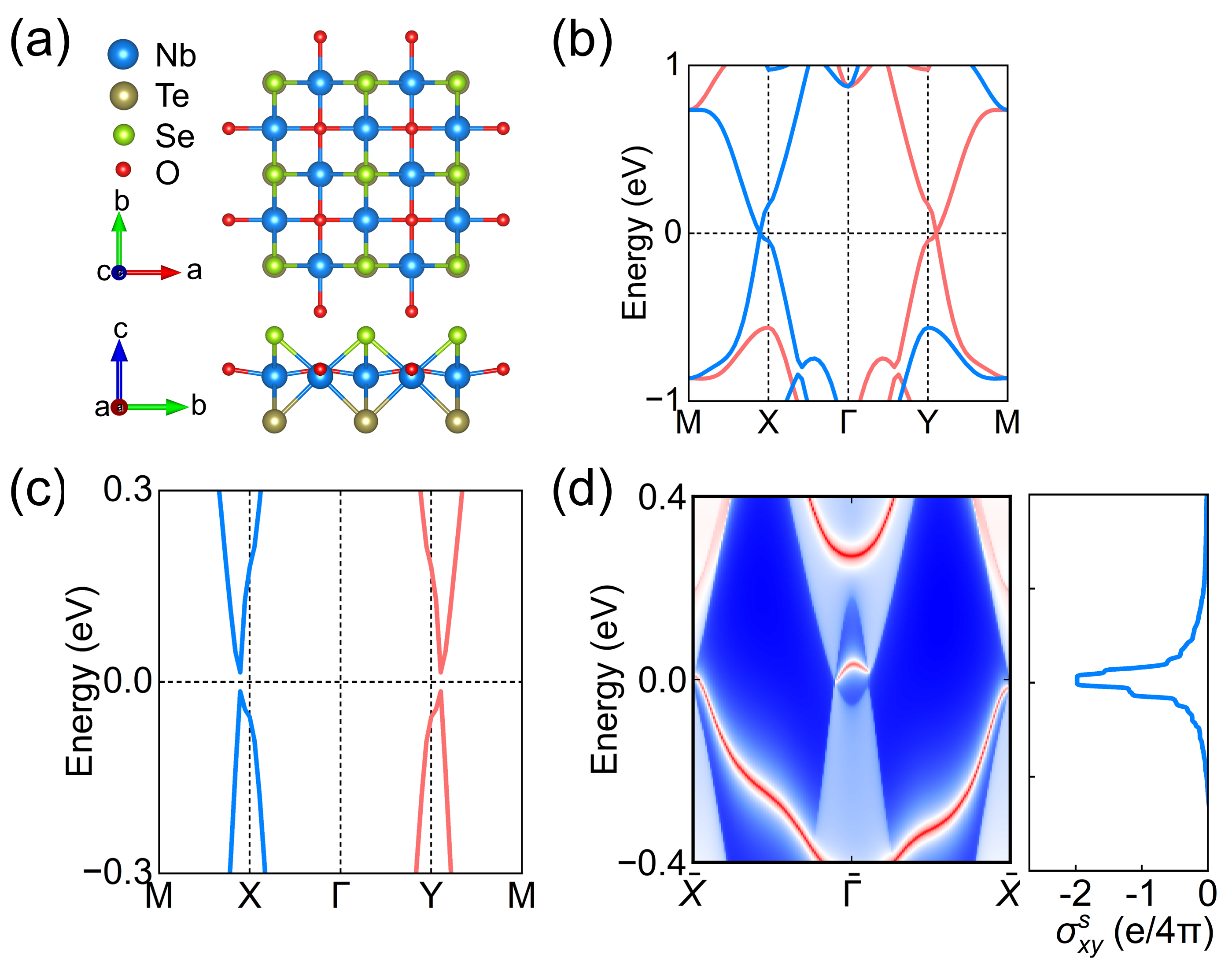} 
\vspace{-1em}%
\caption{(a) Top view and side view of monolayer $\text{Nb}_2\text{SeTeO}$.
(b) The band structure without SOC.
(c) The band structure with SOC. 
(d) The edge states and SHC of monolayer $\text{Nb}_2\text{SeTeO}$  along the [100] direction.}
\label{fig:label3}
\end{figure}

\begin{figure*}[htbp]
\centering
\includegraphics[width=6in]{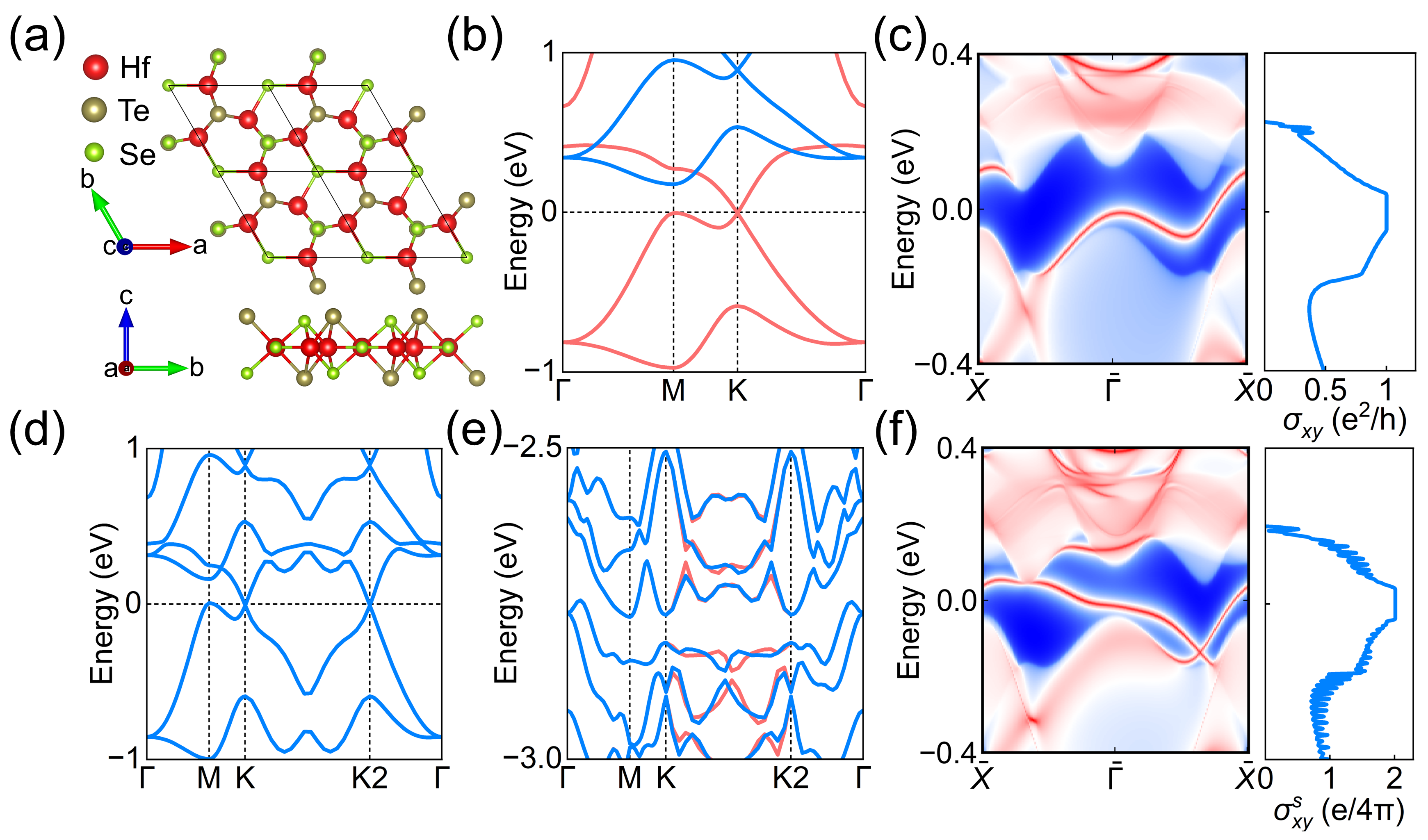} 
\vspace{-1.5em}
\caption{(a) Top view and side view of monolayer $\text{Hf}_3\text{Se}_3\text{Te}_2$.
(b) Band structure without SOC of monolayer $\text{Hf}_3\text{Se}_3\text{Te}_2$.
(c) Edge states and anomalous Hall conductivity (AHC)
 of monolayer $\text{Hf}_3\text{Se}_3\text{Te}_2$. 
(d) Band structure without SOC of bilayer $\text{Hf}_3\text{Se}_3\text{Te}_2$.
(e) Zoom-in band structure  without SOC of bilayer $\text{Hf}_3\text{Se}_3\text{Te}_2$.
(f) The edge states and SHC of bilayer $\text{Hf}_3\text{Se}_3\text{Te}_2$ along the [100] direction.}
\label{fig:label4}
\end{figure*}

\textit{Candidate 1:$\text{Nb}_2\text{SeTeO}$}---Monolayer  $\text{Nb}_2\text{SeTeO}$ is a square structure with layer group of P4mm (No. 55) [see Fig. 3(a)]. Monolayer $\text{Nb}_2\text{Te}_2\text{O}$ is structurally analogous to experimentally synthesized $\text{V}_2\text{Te}_2\text{O}$~\cite{Zhang2025_NatPhys}, and monolayer $\text{Nb}_2\text{SeTeO}$ can be obtained by substituting one Te atom with a Se atom. After optimization, the lattice constants of monolayer  $\text{Nb}_2\text{SeTeO}$ are a = b = 4.12 Å. The dynamical stability are confirmed in Supplementary Material~\cite{SM_Note}. We confirm that the ground state of monolayer $\text{Nb}_2\text{SeTeO}$ exhibits an AM configuration~\cite{SM_Note}, and belongs to MPG 13.3.46, which is exactly a target altermagnetic QSHE candidate. In the absence of SOC, monolayer $\text{Nb}_2\text{SeTeO}$ exhibits non-relativistic spin splitting, with one pair of Weyl points of opposite spin emerging along the $M$-$X$ and $M$-$Y$ high-symmetry paths, respectively [see Fig. 3(b)]. When SOC is considered, band inversion occurs at the Weyl points, generating a pair of valleys with opposite spins along each of the $M$-$X$ and $M$-$Y$ paths and yielding characteristic SVL, which is protected by the \(\mathcal{C}_{4z}\mathcal{T}\) symmetry[see Fig. 3(c)]. Each Weyl point contributes a Chern number of $|C|=1$, which is consistent with our theoretical predictions. The edge state of monolayer $\text{Nb}_2\text{SeTeO}$ reveals two topologically protected edge states with opposite chirality near the $\tilde{\Gamma}$ and $\tilde{X}$ points, as shown in Fig. 3(d). Meanwhile, our calculations demonstrate that the SHC of monolayer $\text{Nb}_2\text{SeTeO}$ is quantized as $|\sigma_{xy}^{s}|=2e/4\pi$. Our calculations results demonstrate that monolayer $\text{Nb}_2\text{SeTeO}$ is an ideal candidate material, which validates our theoretical analysis.

\textit{Candidate 2:$\text{Hf}_3\text{Se}_3\text{Te}_2$}---Monolayer $\text{Hf}_3\text{Se}_3\text{Te}_2$ adopts a hexagonal crystal structure with layer group P321 (No. 68), hosting out-of-plane threefold rotational \(C_{3z}\) and in-plane twofold rotational \(C_{2x}\) symmetries [see Fig. 4(a)]. The optimized lattice constants are a = b = 6.58 
Å. Among all calculated magnetic configurations, the FM phase exhibits the lowest  energy and hence is energetically favored over the AFM arrangements~\cite{SM_Note}.
The spin-resolved band structure of monolayer $\text{Hf}_3\text{Se}_3\text{Te}_2$  without 
SOC presented in Fig. 4(b). SOC opens a global band gap of 203 meV at the Weyl points, identifying monolayer $\text{Hf}_3\text{Se}_3\text{Te}_2$  as a promising candidate for large-gap Chern insulators. Calculations of edge state and AHC further verify the existence of chiral edge states with Chern number $C=1$ [see Fig. 4(c)]. Subsequently, we construct the AA-stacked bilayer of $\text{Hf}_3\text{Se}_3\text{Te}_2$ from its monolayer via the stacking operator $\hat{P}=E$, with $E$ denoting the identity operator, and the resultant system retains the layer group $P321$ (No. 68)~\cite{SM_Note}. In our calculations, we manually set the system of bilayer \(\text{Hf}_3\text{Se}_3\text{Te}_2\) to adopt an interlayer antiferromagnetic configuration. Under the SOC, the bilayer $\text{Hf}_3\text{Se}_3\text{Te}_2$ belongs to MPG 18.1.65, which is exactly a target altermagnetic QSHE candidate. The bilayer hosts a band structure analogous to its monolayer counterpart, featuring a pair of Weyl points near the Fermi level [see Fig. 4(d)]. Distinct from monolayer $\text{Hf}_3\text{Se}_3\text{Te}_2$, the bilayer exhibits nonrelativistic spin splitting originating from AM along the $K$–$K2$ path, which is protected by $C_{2x}$, $C_{2y}$ and $C_{2xy}$ symmetry [see Fig. 4(e)]. Similar to the monolayer system, SOC opens a global band gap at the Weyl points. Owing to the intrinsic altermagnetic nature of the system, band inversion drives the emergence of the QSHE rather than the quantum anomalous Hall effect (QAHE)~\cite{Haldane1988,Chang2023_RMP,Chang2013_Science,Deng2020_Science}, as further validated by edge-state and SHC calculations [see Fig. 4(e)]. Notably, since monolayer $\text{Hf}_3\text{Se}_3\text{Te}_2$ is a Chern insulator, it inherently bears nonvanishing Berry curvature. The symmetries $C_{2x}$, $C_{2y}$, and $C_{2xy}$ simultaneously flip spin, Berry curvature, valley and layer degrees of freedom, endowing the bilayer with characteristic SVLL QSHE. In addition, AB stacking is obtained by translating the upper layer of Hf$_3$ Se$_3$Te$_2$ along the vector $\mathbf{r} = 1/3\mathbf{a} + 2/3\mathbf{b}$, and exhibits properties similar to AA stacking ~\cite{SM_Note}.

\textit{Discussion and conclusion}---Owing to the intrinsic SVL or SVLL characteristics, the altermagnetic QSHE can be readily modulated by external fields. For AMs hosting SVL QSHE, uniaxial strain or staggered sublattice potential drives the transition from QSHE to the QAHE~\cite{Jiang2025,Chen2026_arXiv2}. In contrast, for SVLL QSHE AMs, an external electric field enables the realization of quantum layer Hall effect and quantum layer spin Hall effect~\cite{Feng2025L,Liu2026L}. Furthermore, without breaking any crystalline symmetry, the altermagnetic QSHE can cooperate with 2D second-order topological insulators to realize exotic spin-corner coupling~\cite{Niu2026,Yao2024L,Gong2024_AM}.  This work resolves the critical bottlenecks of conventional QSHE and provides a feasible platform for the experimental realization and practical application of robust low-dissipation topological spin devices.

In conclusion, we establish the relevant symmetry constraints and identify the MPG capable of hosting altermagnetic QSHE. Such systems generally exhibit prominent SVL or SVLL. The cooperation between band inversion and spin-valley locking gives rise to the emergence of helical edge states. We predict two promising candidates for altermagnetic QSHE, namely monolayer $\text{Nb}_2\text{SeTeO}$ and bilayer $\text{Hf}_3\text{Se}_3\text{Te}_2$. This work not only improves the classification framework of magnetic topological insulators, but also delivers feasible theoretical guidance for experimental fabrication of low-dissipation topological spin devices and the controllable manipulation of quantum spin Hall phase transitions.

\textit{Acknowledgments}---This work was financially supported by the Fundamental Research Funds for the Central Universities (No. N25LPY025), the Liao Ning Revitalization Talents Program (Grant No. XLYC1907033), and the Natural Science Foundation of Liaoning province (Grant No. 2023MS-072 and No. 2024-MSBA-36). X. K. acknowledges the start-up funding from Northeastern University, China.

\textit{Data availability}---The data are not publicly available. The data are available from the authors upon reasonable request.

\bibliography{ref} 

@article{Bansil2016,
  author  = {Bansil, A. and Lin, H. and Das, Tanmoy},
  title   = {Colloquium: Topological band theory},
  journal = {Rev. Mod. Phys.},
  year    = {2016},
  volume  = {88},
  pages   = {021004},
  doi     = {10.1103/RevModPhys.88.021004}
}

@article{Hasan2010,
  author  = {Hasan, M. Z. and Kane, C. L.},
  title   = {Colloquium: Topological insulators},
  journal = {Rev. Mod. Phys.},
  year    = {2010},
  volume  = {82},
  pages   = {3045},
  doi     = {10.1103/RevModPhys.82.3045}
}

@article{Qi2011,
  author  = {Qi, Xiao-Liang and Zhang, Shou-Cheng},
  title   = {Topological insulators and superconductors},
  journal = {Rev. Mod. Phys.},
  year    = {2011},
  volume  = {83},
  pages   = {1057},
  doi     = {10.1103/RevModPhys.83.1057}
}

@article{Breunig2022,
  author  = {Breunig, Oliver and Ando, Yoichi},
  title   = {Opportunities in topological insulator devices},
  journal = {Nat. Rev. Phys.},
  year    = {2022},
  volume  = {4},
  pages   = {184},
  doi     = {10.1038/s42254-021-00402-6}
}

@article{Xiao2021,
  author  = {Xiao, Jierui and Yan, Binghai},
  title   = {First-principles calculations for topological quantum materials},
  journal = {Nat. Rev. Phys.},
  year    = {2021},
  volume  = {3},
  pages   = {283},
  doi     = {10.1038/s42254-020-00271-w}
}

@article{Kane2005,
  author  = {Kane, C. L. and Mele, E. J.},
  title   = {${Z}_{2}$ topological order and the quantum spin Hall effect},
  journal = {Phys. Rev. Lett.},
  year    = {2005},
  volume  = {95},
  pages   = {146802},
  doi     = {10.1103/PhysRevLett.95.146802}
}

@article{Kane2005graphene,
  author  = {Kane, C. L. and Mele, E. J.},
  title   = {Quantum spin Hall effect in graphene},
  journal = {Phys. Rev. Lett.},
  year    = {2005},
  volume  = {95},
  pages   = {226801},
  doi     = {10.1103/PhysRevLett.95.226801}
}

@article{Maciejko2009,
  author  = {Maciejko, Joseph and Liu, Chaoxing and Oreg, Yuval and Qi, Xiao-Liang and Wu, Congjun and Zhang, Shou-Cheng},
  title   = {{Kondo} Effect in the Helical Edge Liquid of the Quantum Spin Hall State},
  journal = {Phys. Rev. Lett.},
  year    = {2009},
  volume  = {102},
  pages   = {256803},
  doi     = {10.1103/PhysRevLett.102.256803}
}

@article{Kimme2016,
  author  = {Kimme, Lukas and Rosenow, Bernd and Brataas, Arne},
  title   = {Backscattering in helical edge states from a magnetic impurity and {Rashba} disorder},
  journal = {Phys. Rev. B},
  year    = {2016},
  volume  = {93},
  pages   = {081301},
  doi     = {10.1103/PhysRevB.93.081301}
}

@article{Jack2020,
  author = {J{\"a}ck, Berthold and Xie, Yonglong and Bernevig, B. Andrei and Yazdani, Ali},
  title = {Observation of backscattering induced by magnetism in a topological edge state},
  journal = {Proc. Natl. Acad. Sci. U.S.A.},
  volume = {117},
  number = {28},
  pages = {16214--16218},
  year = {2020},
  doi = {10.1073/pnas.2005071117}
}

@article{Bernevig2006,
  author={Bernevig, B. Andrei and Hughes, Taylor L. and Zhang, Shou-Cheng},
 title   = {Quantum spin {Hall} effect and topological phase transition in {{HgTe}} quantum wells},
  journal={Science},
  volume={314},
  number={5806},
  pages={1757--1761},
  year={2006},
  doi={10.1126/science.1133734}
}

@article{Konig2007,
  author={K{\"o}nig, Markus and Wiedmann, Steffen and Br{\"u}ne, Christoph and Roth, Andreas and Buhmann, Hartmut and Molenkamp, Laurens W. and Qi, Xiao-Liang and Zhang, Shou-Cheng},
  title={Quantum Spin Hall Insulator State in {{HgTe}} Quantum Wells},
  journal={Science},
  volume={318},
  number={5851},
  pages={766--770},
  year={2007},
  doi={10.1126/science.1148047}
}

@article{Tang2017,
  author={Tang, Shujie and Zhang, Chaofan and Wong, Dillon and Pedramrazi, Zahra and Tsai, Hsin-Zon and Jia, Chunjing and Moritz, Brian and Claassen, Martin and Ryu, Hyejin and Kahn, Salman and Jiang, Juan and Yan, Hao and Hashimoto, Makoto and Lu, Donghui and Moore, Robert G. and Hwang, Chan-Cuk and Hwang, Choongyu and Hussain, Zahid and Chen, Yulin and Ugeda, Miguel M. and Liu, Zhi and Xie, Xiaoming and Devereaux, Thomas P. and Crommie, Michael F. and Mo, Sung-Kwan and Shen, Zhi-Xun},
  title   = {Quantum spin {Hall} state in monolayer {${1T}'$-${\mathrm{WTe}}_{2}$}},
  journal={Nat. Phys.},
  volume={13},
  pages={683--687},
  year={2017},
  doi={10.1038/nphys4174}
}

@article{Xu2006,
  author={Xu, Cenke and Moore, J. E.},
  title={Stability of the quantum spin {Hall} effect: Effects of interactions, disorder, and {$Z_2$} topology},
  journal={Phys. Rev. B},
  volume={73},
  pages={045322},
  year={2006},
  doi={10.1103/PhysRevB.73.045322}
}

@article{Hattori2011J,
  author={Hattori, Kiminori},
  title={Quantized Spin Transport in Magnetically-Disordered Quantum Spin Hall Systems},
  journal={J. Phys. Soc. Jpn.},
  volume={80},
  pages={124712},
  year={2011},
  doi={10.1143/jpsj.80.124712}
}

@article{Smejkal2022_PRX,
  author  = {{\v{S}}mejkal, L. and Hellenes, A. B. and Gonz{\'a}lez-Hern{\'a}ndez, R. and Sinova, J. and Jungwirth, T.},
  title   = {Giant and tunneling magnetoresistance in unconventional collinear antiferromagnets with nonrelativistic spin-momentum coupling},
  journal = {Phys. Rev. X},
  year    = {2022},
  volume  = {12},
  pages   = {011028},
  doi     = {10.1103/PhysRevX.12.011028}
}

@article{Smejkal2022_PRX2,
  author  = {{\v{S}}mejkal, L. and Sinova, J. and Jungwirth, T.},
  title   = {Emerging research landscape of altermagnetism},
  journal = {Phys. Rev. X},
  year    = {2022},
  volume  = {12},
  pages   = {040501},
  doi     = {10.1103/PhysRevX.12.040501}
}

@article{Mazin2022,
  author  = {Mazin, I. and {The PRX Editors}},
  title   = {Altermagnetism---A new punch line of fundamental magnetism},
  journal = {Phys. Rev. X},
  year    = {2022},
  volume  = {12},
  pages   = {040002},
  doi     = {10.1103/PhysRevX.12.040002}
}

@article{Mazin2023,
  author  = {Mazin, I.},
  title   = {Altermagnetism in {MnTe}: Origin, predicted manifestations, and routes to detwinning},
  journal = {Phys. Rev. B},
  year    = {2023},
  volume  = {107},
  pages   = {L100418},
  doi     = {10.1103/PhysRevB.107.L100418}
}

@article{GonzalezHernandez2021,
  author  = {Gonz{\'a}lez-Hern{\'a}ndez, R. and {\v{S}}mejkal, L. and V{\'y}born{\'y}, K. and Yahagi, Y. and Sinova, J. and Jungwirth, T. and {\v{Z}}elezn{\'y}, J.},
  title   = {Efficient electrical spin splitter based on nonrelativistic collinear antiferromagnetism},
  journal = {Phys. Rev. Lett.},
  year    = {2021},
  volume  = {126},
  pages   = {127701},
  doi     = {10.1103/PhysRevLett.126.127701}
}

@article{Sinova2020,
  author  = {{\v{S}}mejkal, Libor and Gonz{\'a}lez-Hern{\'a}ndez, Rafael and Jungwirth, T. and Sinova, J.},
  title   = {Crystal time-reversal symmetry breaking and spontaneous {Hall} effect in collinear antiferromagnets},
  journal = {Sci. Adv.},
  volume  = {6},
  number  = {23},
  pages   = {eaaz8809},
  year    = {2020},
  doi     = {10.1126/sciadv.aaz8809}
}

@article{Dev2025,
  author  = {Sheoran, Sajjan and Dev, Pratibha},
  title   = {Spontaneous anomalous {Hall} effect in two-dimensional altermagnets},
  journal = {Phys. Rev. B},
  year    = {2025},
  volume  = {111},
  pages   = {184407},
  doi     = {10.1103/PhysRevB.111.184407}
}

@article{Ma2021,
  author  = {Ma, Hai-Yuan and Hu, Min and Li, Na and Liu, Jian and Yao, Wang and Yan, Binghai},
  title   = {Multifunctional antiferromagnetic materials with giant piezomagnetism and noncollinear spin current},
  journal = {Nat. Commun.},
  year    = {2021},
  volume  = {12},
  pages   = {2846},
  doi     = {10.1038/s41467-021-23127-7}
}

@article{Tomas2022,
  author = {{\v{S}}mejkal, Libor and Hellenes, Anna Birk and Gonz{\'a}lez-Hern{\'a}ndez, Rafael and Sinova, Jairo and Jungwirth, Tomas},
  title = {Giant and Tunneling Magnetoresistance in Unconventional Collinear Antiferromagnets with Nonrelativistic Spin-Momentum Coupling},
  journal = {Phys. Rev. X},
  volume = {12},
  pages = {011028},
  year = {2022},
  doi = {10.1103/PhysRevX.12.011028}
}

@article{Venderbos2025,
  author = {Antonenko, Daniil S. and Fernandes, Rafael M. and Venderbos, J{\"o}rn W. F.},
  title = {Mirror Chern Bands and Weyl Nodal Loops in Altermagnets},
  journal = {Phys. Rev. Lett.},
  volume = {134},
  pages = {096703},
  year = {2025},
  doi = {10.1103/PhysRevLett.134.096703}
}

@article{Niu2025l,
  author  = {Zhang, Zequn and Bai, Yingxi and Zou, Xiaorong and Huang, Baibiao and Dai, Ying and Niu, Chengwang},
  title   = {Altermagnetic quantum spin {Hall} effect in a {Chern} homobilayer},
  journal = {Phys. Rev. B},
  volume  = {112},
  pages   = {085128},
  year    = {2025},
  doi = {10.1103/zbbr-426l}
}

@article{Jiang2025,
  author  = {Jiang, Y. Q. and Zhang, X. G. and Bai, H. Y. and Tian, Y. P. and Zhang, B. Y. and Gong, W. J. and Kong, X. R.},
  title   = {Strain-engineering spin-valley locking effect in altermagnetic monolayer with multipiezo properties},
  journal = {Appl. Phys. Lett.},
  year    = {2025},
  volume  = {126},
  pages   = {053102},
  doi     = {https://doi.org/10.1063/5.0252374}
}

@article{Yao2026,
  author  = {Zhang, Run-Wu and Cui, Chaoxi and Wang, Yang and Duan, Jingyi and Yu, Zhi-Ming and Yao, Yugui},
  title   = {Quantized spin {Hall} conductivity in altermagnetic $\mathrm{Fe}_2\mathrm{Te}_2\mathrm{O}$ with mirror-spin coupling},
  journal = {Phys. Rev. B},
  volume  = {113},
  pages   = {L161115},
  year    = {2026},
  doi = {10.1103/s9mm-5662},
}

@misc{Chen2026_arXiv2,
      title={Altermagnets Enable Gate-Switchable Helical and Chiral Topological Transport with Spin-Valley-Momentum-Locked Dual Protection}, 
      author={Xianzhang Chen and Jiayong Zhang and Bowen Hao and Jiahui Qian and Ziye Zhu and Igor Zutic and Zhenyu Zhang and Tong Zhou},
      year={2026},
      eprint={2603.06487},
      archivePrefix={arXiv},
}

@article{Bernardo2025,
  title = {Spin Chern number in altermagnets},
  author = {Gonz\'alez-Hern\'andez, Rafael and Serrano, Higinio and Uribe, Bernardo},
  journal = {Phys. Rev. B},
  volume = {111},
  issue = {8},
  pages = {085127},
  numpages = {11},
  year = {2025},
  month = {Feb},
  publisher = {American Physical Society},
  doi = {10.1103/PhysRevB.111.085127},
  url = {https://link.aps.org/doi/10.1103/PhysRevB.111.085127}
}

@misc{Feng2025_arXiv,
      title={Multiple Topological Phases Controlled via Strain in Two-Dimensional Altermagnets}, 
      author={Zesen Fu and Mengli Hu and Aolin Li and Haiming Duan and Junwei Liu and Fangping Ouyang},
      year={2025},
      eprint={2507.22474},
      archivePrefix={arXiv}
}

@article{TAN20262196,
  author = {Chao-Yang Tan and Panjun Feng and Ze-Feng Gao and Fengjie Ma and Peng-Jie Guo and Zhong-Yi Lu},
  title = {Stacking-induced type-{II} quantum spin Hall insulators with high spin Chern number in unconventional magnetism},
  journal = {Science Bulletin},
  volume = {71},
  pages = {2196--2199},
  year = {2026},
  doi = {10.1016/j.scib.2026.04.018}
}

@article{Chen2026,
  author  = {Chen, Zhiyu and Zhan, Fangyang and Qin, Zheng and Ma, Da-Shuai and Xu, Dong-Hui and Wang, Rui},
  title   = {Quantum Spin {Hall} Effect with Extended Topologically Protected Features in Altermagnetic Multilayers},
  journal = {Nano Lett.},
  year    = {2026},
  volume  = {26},
  number  = {12},
  pages   = {4197--4203},
  doi     ={https://doi.org/10.1021/acs.nanolett.6c00136}
}

@article{Liu2026,
  author    = {Yu, Jiangtao and Bai, Jingbo and Yang, Yali and Qian, Shifeng and Wang, Xiaotian and Liu, Zhuhong},
  title     = {Diverse Landscape of Tunable Magnetic, Topological, and Ferroelectric States in {2D} $\mathrm{Ti}_3\mathrm{Se}_3\mathrm{Te}_2$},
  journal   = {Adv. Sci.},
  volume    = {13},
  number    = {11},
  pages     = {2524385},
  year      = {2026},
  doi       = {10.1002/advs.202524385}
}

@article{Zou2025,
  author  = {Zou, X. R. and Feng, X. R. and Dai, Y. and Huang, B. B. and Niu, C. W.},
  title   = {{Floquet} quantum anomalous {Hall} effect with in-plane magnetization in two-dimensional altermagnets},
  journal = {ACS Nano},
  year    = {2025},
  volume  = {19},
  pages   = {35575--35580},
  doi     = {https://doi.org/10.1021/acsnano.5c10277}
}

@article{Zhang2024,
  author  = {Zhang, R. W. and Cui, C. X. and Li, R. Z. and Duan, J. Y. and Li, L. and Yu, Z. M. and Yao, Y. G.},
  title   = {Predictable gate-field control of spin in altermagnets with spin-layer coupling},
  journal = {Phys. Rev. Lett.},
  year    = {2024},
  volume  = {133},
  pages   = {056401},
  doi     = {10.1103/PhysRevLett.133.056401}
}

@article{Yang2025,
  author  = {Yang, N. J. and Huang, Z. G. and Zhang, J. M.},
  title   = {Spin-selective second-order topological insulators enabling cornertronics in two-dimensional altermagnets},
  journal = {Nano Lett.},
  year    = {2025},
  volume  = {25},
  pages   = {15495--15500},
  doi     = {https://doi.org/10.1021/acs.nanolett.5c03191}
}

@article{Liu2025,
  author  = {Liu, K. H. and Zhao, M. W.},
  title   = {Altermagnetism and higher-order topological states in bilayer {Chern} insulators},
  journal = {Phys. Rev. B},
  year    = {2025},
  volume  = {112},
  pages   = {L241405},
  doi = {10.1103/xyjn-8dqk}
}

@article{Hu2026,
  author  = {Wang, Qianjun and Wu, Ruqian and Hu, Jun},
  title   = {Spin-biased quantum spin {Hall} effect in altermagnetic {Lieb} lattice},
  journal = {Phys. Rev. B},
  year    = {2026},
  volume  = {113},
  pages   = {L161101},
  doi     = {10.1103/kqwx-v6jv}
}

@article{Jing2026_CJP,
  author  = {Jing, Tao and Onyx, Terchie-Duku and Liang, Dongmei and Xiong, Yongchen and Hu, Yongjin and Deng, Mingsen},
  title   = {Quantum spin {Hall} insulator with altermagnetism in $\mathrm{NiNbSe}_2$ bilayer},
  journal = {Chin. J. Phys.},
  year    = {2026},
  volume  = {102},
  pages   = {141--149},
  doi     = {10.1016/j.cjph.2026.03.033}
}

@article{Feng2024,
  author  = {Ma, Hai-Young and Jia, Jin-Feng},
  title   = {Altermagnetic topological insulator and the selection rules},
  journal = {Phys. Rev. B},
  year    = {2024},
  volume  = {110},
  pages   = {064426},
  doi     = {10.1103/PhysRevB.110.064426}
}

@misc{Fu2025_arXiv,
      title={Multiple Topological Phases Controlled via Strain in Two-Dimensional Altermagnets}, 
      author={Zesen Fu and Mengli Hu and Aolin Li and Haiming Duan and Junwei Liu and Fangping Ouyang},
      year={2025},
      eprint={2507.22474},
      archivePrefix={arXiv},
}

@article{Ma2021_JPCM,
  author  = {Ma, Hai-Yuan and Guan, Dandan and Wang, Shiyong and Li, Yaoyi and Liu, Canhua and Zheng, Hao and Jia, Jin-Feng},
  title   = {Quantum spin {Hall} and quantum anomalous {Hall} states in magnetic $\mathrm{Ti}_2\mathrm{Te}_2\mathrm{O}$ single layer},
  journal = {J. Phys.: Condens. Matter},
  year    = {2021},
  volume  = {33},
  pages   = {21LT01},
  doi     = {10.1088/1361-648X/abe647}
}

@article{Zhang2009,
  author  = {Zhang, Haijun Likelihood and Liu, Chao-Xing and Qi, Xiao-Liang and Dai, Xi and Fang, Zhong and Zhang, Shou-Cheng},
  title   = {Topological insulators in $\mathrm{Bi}_2\mathrm{Se}_3$, $\mathrm{Bi}_2\mathrm{Te}_3$ and $\mathrm{Sb}_2\mathrm{Te}_3$ with a single {Dirac} cone on the surface},
  journal = {Nat. Phys.},
  year    = {2009},
  volume  = {5},
  number  = {6},
  pages   = {438--442},
  doi     = {10.1038/nphys1270}
}

@article{Chen2024,
  author  = {Chen, X. B. and Ren, J. and Zhu, Y. Z. and Yu, Y. T. and Zhang, A. and Liu, P. F. and Li, J. Y. and Liu, Y. T. and Li, C. H. and Liu, Q. H.},
  title   = {Enumeration and Representation Theory of Spin Space Groups},
  journal = {Phys. Rev. X},
  year    = {2024},
  volume  = {14},
  pages   = {031038},
  doi     = {10.1103/PhysRevX.14.031038}
}

@article{Xiao2024,
  author  = {Xiao, Z. Y. and Zhao, J. Z. and Li, Y. Q. and Shindou, R. and Song, Z. D.},
  title   = {Spin Space Groups: Full Classification and Applications},
  journal = {Phys. Rev. X},
  year    = {2024},
  volume  = {14},
  pages   = {031037},
  doi     = {10.1103/PhysRevX.14.031037}
}

@article{Jiang2024,
  author  = {Jiang, Y. and Song, Z. Y. and Zhu, T. N. and Fang, Z. and Weng, H. M. and Liu, Z. X. and Yang, J. and Fang, C.},
  title   = {Enumeration of Spin-Space Groups: Toward a Complete Description of Symmetries of Magnetic Orders},
  journal = {Phys. Rev. X},
  year    = {2024},
  volume  = {14},
  pages   = {031039},
  doi     = {10.1103/PhysRevX.14.031039}
}

@article{Elcoro2021,
  author  = {Elcoro, L. and Wieder, B. J. and Song, Z. and Regnault, N. and Bradlyn, B. and Bernevig, B. A.},
  title   = {Magnetic topological quantum chemistry},
  journal = {Nat. Commun.},
  year    = {2021},
  volume  = {12},
  pages   = {5965},
  doi     = {10.1038/s41467-021-26241-8}
}

@article{Zhang2022_CPC,
  author  = {Zhang, Zeying and Yu, Zhi-Ming and Liu, Gui-Bin and Yao, Yugui},
  title   = {{MagneticTB}: A package for tight-binding model of magnetic and nonmagnetic materials},
  journal = {Comput. Phys. Commun.},
  year    = {2022},
  volume  = {270},
  pages   = {108153},
  doi     = {10.1016/j.cpc.2021.108153}
}

@article{Zhang2025_NatPhys,
  author  = {Zhang, F. and Cheng, X. and Yin, Z. and Han, S. and Chen, X. B. and Zhou, P. and Jiang, Z. and Liu, C. and Liu, Q. H.},
  title   = {Crystal-symmetry-paired spin--valley locking in a layered room-temperature metallic altermagnet candidate},
  journal = {Nat. Phys.},
  year    = {2025},
  volume  = {21},
  pages   = {760--767},
  doi     = {10.1038/s41567-025-02864-2}
}

@misc{SM_Note,
  key = {},
  note = {See Supplemental Material for the MPGs and  MLGs of the altermagnetic QSHE, effective tight-binding Hamiltonian, Brillouin Zones for models and materials in the Main Text, computational methods,  other candidate materials, supplemental figures .}
}

@article{Haldane1988,
  author  = {Haldane, F. D. M.},
  title   = {Model for a quantum {Hall} effect without {Landau} levels: Condensed-matter realization of the ``parity anomaly''},
  journal = {Phys. Rev. Lett.},
  year    = {1988},
  volume  = {61},
  pages   = {2015--2018},
  doi     = {10.1103/PhysRevLett.61.2015}
}

@article{Chang2023_RMP,
  author  = {Chang, Cui-Zu and Liu, Chao-Xing and MacDonald, Allan H.},
  title   = {Colloquium: Quantum anomalous {Hall} effect},
  journal = {Rev. Mod. Phys.},
  year    = {2023},
  volume  = {95},
  pages   = {011002},
  doi     = {10.1103/RevModPhys.95.011002}
}

@article{Chang2013_Science,
  author  = {Chang, Cui-Zu and Zhang, Jinsong Xia and Feng, Xiao and Shen, Jie and Zhang, Zuocheng and Guo, Minghua and Li, Kang and Ou, Yunbo and Wei, Pang and Wang, Li-Li and Ji, Zhong-Qing and Feng, Yang and Ji, Shuaihua and Chen, Xi and Jia, Jinfeng and Dai, Xi and Fang, Zhong and Zhang, Shou-Cheng and He, Ke and Wang, Yayu and Lu, Li and Ma, Xu-Cun and Xue, Qi-Kun},
  title   = {Experimental observation of the quantum anomalous {Hall} effect in a magnetic topological insulator},
  journal = {Science},
  year    = {2013},
  volume  = {340},
  number  = {6129},
  pages   = {167--170},
  doi     = {10.1126/science.1234414}
}

@article{Deng2020_Science,
  author  = {Deng, Yujun and Yu, Yijun and Shi, Meng Zhu and Guo, Zhiandong and Xu, Zihan and Wang, Jing and Chen, Xian Hui and Zhang, Yuanbo},
  title   = {Quantum anomalous {Hall} effect in intrinsic magnetic topological insulator $\mathrm{MnBi}_2\mathrm{Te}_4$},
  journal = {Science},
  year    = {2020},
  volume  = {367},
  number  = {6480},
  pages   = {895--900},
  doi     = {10.1126/science.aax8156}
}

@article{Feng2025L,
  title = {Layer Hall and layer spin Hall effects in two-dimensional altermagnets induced by spin-layer coupling},
  author = {Wang, Xiangju and Liu, Siyuan and Bai, Ling and Zhang, Run-Wu and Yao, Yugui and Feng, Wanxiang},
  journal = {Phys. Rev. B},
  volume = {112},
  issue = {13},
  pages = {134421},
  numpages = {8},
  year = {2025},
  month = {Oct},
  publisher = {American Physical Society},
  doi = {10.1103/643d-wkc1},
  url = {https://link.aps.org/doi/10.1103/643d-wkc1}
}

@article{Liu2026L,
  title = {Electric field control of spin-valley polarization and the spin Hall effect in altermagnets},
  author = {Zhou, Hongfu and Wang, Xinyi and Liu, Fujun},
  journal = {Phys. Rev. B},
  volume = {113},
  issue = {9},
  pages = {094427},
  numpages = {9},
  year = {2026},
  month = {Mar},
  publisher = {American Physical Society},
  doi = {10.1103/tlpw-95yc},
  url = {https://link.aps.org/doi/10.1103/tlpw-95yc}
}

@article{Niu2026,
  title = {Topological control of corner and edge states in an altermagnetic {$\mathrm{Fe}_{2}\mathrm{Se}_{2}\mathrm{O}$} monolayer},
  author = {Zhang, Yilin and Chen, Zhiqi and Zou, Xiaorong and Huang, Baibiao and Dai, Ying and Niu, Chengwang},
  journal = {Phys. Rev. B},
  volume = {113},
  issue = {15},
  pages = {155428},
  numpages = {6},
  year = {2026},
  month = {Apr},
  publisher = {American Physical Society},
  doi = {10.1103/bkhw-bxgp},
  url = {https://link.aps.org/doi/10.1103/bkhw-bxgp}
}

@article{Yao2024L,
  title = {Cornertronics in Two-Dimensional Second-Order Topological Insulators},
  author = {Han, Yilin and Cui, Chaoxi and Li, Xiao-Ping and Zhang, Ting-Ting and Zhang, Zeying and Yu, Zhi-Ming and Yao, Yugui},
  journal = {Phys. Rev. Lett.},
  volume = {133},
  issue = {17},
  pages = {176602},
  numpages = {8},
  year = {2024},
  month = {Oct},
  publisher = {American Physical Society},
  doi = {10.1103/PhysRevLett.133.176602},
  url = {https://link.aps.org/doi/10.1103/PhysRevLett.133.176602}
}

@article{Gong2024_AM,
  author  = {Gong, Jialin and Wang, Yang and Han, Yilin and Cheng, Zhenxiang and Wang, Xiaotian and Yu, Zhi-Ming and Yao, Yugui},
  title   = {Hidden Real Topology and Unusual Magnetoelectric Responses in Two-Dimensional Antiferromagnets},
  journal = {Adv. Mater.},
  year    = {2024},
  volume  = {36},
  number  = {29},
  pages   = {2402232},
  doi     = {10.1002/adma.202402232}
}

@article{smejkal2022B,
  title={Beyond Conventional Ferromagnetism and Antiferromagnetism: A Phase with Nonrelativistic Spin and Crystal Rotation Symmetry},
  author={{\v{S}}mejkal, L. and Sinova, J. and Jungwirth, T.},
  journal={Phys. Rev. X},
  volume={12},
  pages={031042},
  year={2022},
  doi={10.1103/PhysRevX.12.031042}
}

@misc{mazin20231,
 title={Induced monolayer altermagnetism in $\mathrm{MnP(S,Se)_3}$ and $\mathrm{FeSe}$},
      author={Igor Mazin and Rafael González-Hernández and Libor Šmejkal},
      year={2023},
      eprint={2309.02355},
      archivePrefix={arXiv},
}

@article{guo2023,
  title={Quantum anomalous Hall effect in collinear antiferromagnetism},
  author={Guo, P.-J. and Liu, Z.-X. and Lu, Z.-Y.},
  journal={npj Comput. Mater.},
  volume={9},
  pages={70},
  year={2023},
  doi={https://doi.org/10.1038/s41524-023-01025-4},
  url={https://www.nature.com/articles/s41524-023-01154-w}
}

@article{tan2025,
  title={Crystal valley Hall effect},
  author={Tan, C.-Y. and Gao, Z.-F. and Yang, H.-C. and Liu, Z.-X. and Liu, K. and Guo, P.-J. and Lu, Z.-Y.},
  journal={Phys. Rev. B},
  volume={111},
  pages={094411},
  year={2025},
  doi={10.1103/PhysRevB.111.094411},
  url={https://doi.org/10.1103/PhysRevB.111.094411}
}

@article{shi2026,
  title={Tunable quantum layer spin Hall effect in bilayer altermagnetic $\mathrm{Nb}_2\mathrm{Se}\mathrm{Te}\mathrm{O}$},
  author={Shi, H. and Jiang, Y. and Tian, Y. and Wang, W. and Li, S. and Gong, W.-J. and Kong, X.},
  journal={Appl. Phys. Lett.},
  volume={128},
  pages={063101},
  year={2026},
  doi={https://doi.org/10.1063/5.0312073}
}

@article{xu2025L,
    author = {Xu, Runzhang and Gao, Yifan and Liu, Junwei},
    title = {Chemical design of monolayer altermagnets},
    journal = {Nat. Sci. Rev.},
    volume = {13},
    number = {2},
    pages = {nwaf528},
    year = {2026},
    doi={10.1093/nsr/nwaf528}
}

@article{2022MSG,
  author = {Campbell, B. J. and Stokes, H. T. and Perez-Mato, J. M. and Rodr\'iguez-Carvajal, J.},
  title = {Introducing a unified magnetic space-group symbol},
  journal = {Acta Crystallographica Section A},
  year = {2022},
  volume = {A78},
  pages = {99--106},
  doi = {10.1107/S2053273321012912}
}

@article{Ma1,
  title = {Switchable anomalous valley Hall effect in 2D antiferromagnetic system},
  author = {Hui, Jiaxuan and Ma, Yandong and Dai, Ying and Huang, Baibiao and Li, Xinru},
  journal = {Phys. Rev. Mater.},
  volume = {10},
  issue = {2},
  pages = {024402},
  numpages = {8},
  year = {2026},
  month = {Feb},
  publisher = {American Physical Society},
  doi = {10.1103/bdfy-8ff7},
  url = {https://link.aps.org/doi/10.1103/bdfy-8ff7}
}

@article{Ma2,
  author  = {Lei, Chengan and Qian, Zhao and Ma, Yandong and Ahuja, Rajeev},
  title   = {Intrinsic ferroelastic valleytronics in {2D} {Pd4X3Te3} ({X} = {S}, {Se}) materials: {A} new platform for ultrafast intervalley carrier dynamics},
  journal = {Mater. Horiz.},
  volume  = {12},
  number  = {16},
  pages   = {6271--6282},
  year    = {2025},
  doi     = {10.1039/D5MH00567A}
}

@article{Ma3,
  author  = {Wu, Guoli and Feng, Yangyang and Dai, Ying and Huang, Baibiao and Ma, Yandong},
  title   = {Pseudo lattice-breathing driven valley switching in {2D} ferromagnetic lattices},
  journal = {Mater. Horiz.},
  volume  = {13},
  number  = {6},
  pages   = {2911--2917},
  year    = {2026},
  doi     = {10.1039/D5MH02014J}
}

@article{Ma4,
  author    = {Chai, Shuyan and Zhao, Jiangyu and Li, Xinru and Dai, Ying and Huang, Baibiao and Ma, Yandong},
  title     = {van Hove Singularity-Induced Non-Equilibrium Anomalous Valley Hall Effect in a Two-Dimensional Lattice},
  journal   = {Nano Lett.},
  volume    = {25},
  number    = {10},
  pages     = {4108--4114},
  year      = {2025},
  doi       = {10.1021/acs.nanolett.5c00612}
}

@article{Niu2020,
  title = {Antiferromagnetic Topological Insulator with Nonsymmorphic Protection in Two Dimensions},
  author = {Niu, Chengwang and Wang, Hao and Mao, Ning and Huang, Baibiao and Mokrousov, Yuriy and Dai, Ying},
  journal = {Phys. Rev. Lett.},
  volume = {124},
  issue = {6},
  pages = {066401},
  numpages = {6},
  year = {2020},
  month = {Feb},
  publisher = {American Physical Society},
  doi = {10.1103/PhysRevLett.124.066401},
  url = {https://link.aps.org/doi/10.1103/PhysRevLett.124.066401}
}

@article{Jiang20251,
  title = {Intrinsic antiferromagnetic topological insulator and axion state in $\mathrm{V_2WS_4}$},
  author = {Jiang, Yadong and Wang, Huan and Bao, Kejie and Wang, Jing},
  journal = {Phys. Rev. B},
  volume = {111},
  issue = {16},
  pages = {165109},
  numpages = {8},
  year = {2025},
  month = {Apr},
  publisher = {American Physical Society},
  doi = {10.1103/PhysRevB.111.165109},
  url = {https://link.aps.org/doi/10.1103/PhysRevB.111.165109}
}

@article{zou2024,
  author  = {Zou, Xiaorong and Li, Runhan and Chen, Zhiqi and Dai, Ying and Huang, Baibiao and Niu, Chengwang},
  title   = {Engineering Gapless Edge States from Antiferromagnetic Chern Homobilayer},
  journal = {Nano Lett.},
  year    = {2024},
  volume  = {24},
  number  = {1},
  pages   = {450--457},
  doi     = {10.1021/acs.nanolett.3c04304}
}

@article{Kresse1999,
  author = {Kresse, G. and Joubert, D.},
  title = {From ultrasoft pseudopotentials to the projector augmented-wave method},
  journal = {Phys. Rev. B},
  volume = {59},
  pages = {1758},
  year = {1999},
  doi = {10.1103/PhysRevB.59.1758}
}

@article{Kresse1993,
  author = {Kresse, G. and Hafner, J.},
  title = {Ab initio molecular dynamics for liquid metals},
  journal = {Phys. Rev. B},
  volume = {47},
  pages = {558},
  year = {1993},
  doi = {10.1103/PhysRevB.47.558}
}

@article{Kresse1996,
  author = {Kresse, G. and Furthm\"{u}ller, J.},
  title = {Efficient iterative schemes for ab initio total-energy calculations using a plane-wave basis set},
  journal = {Phys. Rev. B},
  volume = {54},
  pages = {11169},
  year = {1996},
  doi = {10.1103/PhysRevB.54.11169}
}

@article{Hohenberg1964,
  author = {Hohenberg, P. and Kohn, W.},
  title = {Inhomogeneous electron gas},
  journal = {Phys. Rev.},
  volume = {136},
  pages = {B864},
  year = {1964},
  doi = {10.1103/PhysRev.136.B864}
}

@article{Kohn1965,
  author = {Kohn, W. and Sham, L. J.},
  title = {Self-consistent equations including exchange and correlation effects},
  journal = {Phys. Rev.},
  volume = {140},
  pages = {A1133},
  year = {1965},
  doi = {10.1103/PhysRev.140.A1133}
}

@article{Ceperley1980,
  author = {Ceperley, D. M. and Alder, B. J.},
  title = {Ground state of the electron gas by a stochastic method},
  journal = {Phys. Rev. Lett.},
  volume = {45},
  pages = {566},
  year = {1980},
  doi = {10.1103/PhysRevLett.45.566}
}

@article{Perdew1981,
  author = {Perdew, J. P. and Zunger, A.},
  title = {Self-interaction correction to density-functional approximations for many-electron systems},
  journal = {Phys. Rev. B},
  volume = {23},
  pages = {5048},
  year = {1981},
  doi = {10.1103/PhysRevB.23.5048}
}

@article{Perdew1996,
  author = {Perdew, J. P. and Burke, K. and Ernzerhof, M.},
  title = {Generalized gradient approximation made simple},
  journal = {Phys. Rev. Lett.},
  volume = {77},
  pages = {3865},
  year = {1996},
  doi = {10.1103/PhysRevLett.77.3865}
}

@article{Monkhorst1976,
  author = {Monkhorst, H. J. and Pack, J. D.},
  title = {Special points for Brillouin zone integrations},
  journal = {Phys. Rev. B},
  volume = {13},
  pages = {5188},
  year = {1976},
  doi = {10.1103/PhysRevB.13.5188}
}

@article{Anasori2016,
  author = {Anasori, B. and Shi, C. and Moon, E. J. and Xie, Y. and Voigt, C. A. and Kent, P. R. C. and May, S. J. and Billinge, S. J. L. and Barsoum, M. W. and Gogotsi, Y.},
  title = {Control of electronic properties of 2D carbides (MXenes) by manipulating their transition metal layers},
  journal = {Nanoscale Horiz.},
  volume = {1},
  pages = {227},
  year = {2016},
  doi = {10.1039/c5nh00125k}
}

@article{Zhang2018,
  author = {Zhang, X. and Wei, J. and Li, R. and Zhang, C. and Zhang, H. and Han, P. and Fan, C.},
  title = {{DFT}+{U} predictions: Structural stability, electronic and optical properties, oxidation activity of BiOCl photocatalysts with 3d transition metals doping},
  journal = {J. Mater. Sci.},
  volume = {53},
  pages = {4494},
  year = {2018},
  doi = {10.1007/s10853-017-1865-0}
}

@article{Yao2022,
  author = {Yao, X. and Ji, J. and Lin, Y. and Sun, Y. and Wang, L. and He, A. and Wang, B. and Lu, P. and He, M. and Zhang, X.},
  title = {$\mathrm{TMB_2C}$ ($\mathrm{TM = Ti,\ V}$): 2D transition metal borocarbide monolayer with intriguing electronic, magnetic and electrochemical properties},
  journal = {Appl. Surf. Sci.},
  volume = {605},
  pages = {154692},
  year = {2022},
  doi = {10.1016/j.apsusc.2022.154692}
}

@article{Togo2015,
  author = {Togo, A. and Tanaka, I.},
  title = {First principles phonon calculations in materials science},
  journal = {Scr. Mater.},
  volume = {108},
  pages = {1},
  year = {2015},
  doi = {10.1016/j.scriptamat.2015.07.021}
}

@article{Mostofi2008,
  author = {Mostofi, A. A. and Yates, J. R. and Lee, Y. S. and Souza, I. and Vanderbilt, D. and Marzari, N.},
  title = {Wannier90: A tool for obtaining maximally-localised Wannier functions},
  journal = {Comput. Phys. Commun.},
  volume = {178},
  pages = {685},
  year = {2008},
  doi = {10.1016/j.cpc.2007.11.016}
}

@article{Wu2018,
  author = {Wu, Q. and Zhang, S. and Song, H.-F. and Troyer, M. and Soluyanov, A. A.},
  title = {WannierTools: An open-source software package for novel topological materials},
  journal = {Comput. Phys. Commun.},
  volume = {224},
  pages = {405},
  year = {2018},
  doi = {10.1016/j.cpc.2017.09.033}
}

@article{bai2026pt,
  title={PT-Symmetric Antiferromagnets as Building Blocks for Anomalous Transport},
  author={Bai, Ling and Liu, Siyuan and Wang, Xiangju and {\v{S}}mejkal, Libor and Sinova, Jairo and Mokrousov, Yuriy and Yao, Yugui and Feng, Wanxiang},
  journal={Nano Lett.},
  volume={26},
  number={11},
  pages={3934--3939},
  year={2026},
  month={3},
  publisher={American Chemical Society},
  doi={10.1021/acs.nanolett.6c00290}
}
\end{document}


\title{Supplemental Material: Symmetry enforced quantum spin Hall effect in Altermagnets}

\author{Fanzheng Chen}
\affiliation{College of Sciences, Northeastern University, Shenyang 110819, China}

\author{Lixin Zhang}
\affiliation{School of Physics and Optoelectronics, Shandong Normal University, Jinan, 250358, China}

\author{Shuaishuai Niu}
\affiliation{College of Sciences, Northeastern University, Shenyang 110819, China}

\author{Junfeng Ren}
\affiliation{School of Physics and Optoelectronics, Shandong Normal University, Jinan, 250358, China}

\author{Weijiang Gong}
\affiliation{College of Sciences, Northeastern University, Shenyang 110819, China}

\author{Xiangru Kong}
\email{Contact author: kongxiangru@neu.edu.cn}
\affiliation{College of Sciences, Northeastern University, Shenyang 110819, China}

\maketitle

\section{MPGs and MLGs of the altermagnetic QSHE}

Magnetic point groups (MPGs) can be classified into three distinct types.
Type I MPGs, as denoted by \(M_{\text{P1}} = G\), exclude time-reversal symmetry \(\mathcal{T}\) and coincide with conventional crystallographic point groups.
Type II MPGs, written as \(M_{\text{P2}} = G + \mathcal{T}G\), describe nonmagnetic systems, where every symmetry operation of the base point group G is paired with the time-reversal operator \(\mathcal{T}\).
Type III MPGs take the form \(M_{\text{P3}} = H + \mathcal{T}(G-H)\) and originate from a halving H of G~\cite{2022MSG}.

In two-dimensional magnetic systems, all MPG symmetries can be divided into two subsets. The set $\mathcal{S} = \{E,P,C_{nz},C_{2\parallel}T,M_{\parallel}T,\,M_{z},\,S_{nz}\}$ ($n=2, 3, 4, 6$) contains symmetry operations that permit a nonzero $\boldsymbol{\sigma_{xy}}$, whereas the other subset $\mathcal{ST} = \{T,PT,C_{nz}T,C_{2\parallel},M_{\parallel},\\M_{z}T,\,S_{nz}T\}$ consists of operations that forbid $\boldsymbol{\sigma_{xy}}$~\cite{bai2026pt}. For altermagnetic materials, Type II MPGs can be neglected since they describe nonmagnetic systems. In addition, MPGs with $\mathcal{PT}$ symmetry are also excluded, as they correspond to $\mathcal{PT}$-type antiferromagnets. According to the symmetry criteria given in the main text, we screened all Type I MPGs and Type III MPGs without $\mathcal{PT}$ symmetry, and finally identified all qualified MPGs and their corresponding magnetic layer groups (MLGs) that can host the altermagnetic quantum spin Hall effect (QSHE). All results are summarized in Table S1.

To offer a clearer understanding of the symmetry requirement of altermagnetic QSHE, here we consider two concrete examples.

(i) Example I: MPG 11.3.37 ($4'/m$). The symmetry operations of MPG 11.3.37 are generated by the set $\{E,\, P,\, C_{2z},\, C_{4z}T\}$. The BZ of systems possessing this MPG exhibits a square geometry, and its two distinct valleys reside at the high-symmetry $X(\pi,0)$ and $Y(0,\pi)$ points. This valley positioning follows directly from the transformation rules: $PX=P$, $C_{2z}X=X$, and $C_{4z}T\,X=Y$.
The Berry curvature $\Omega_{xy}$ transforms under these symmetries as $P\Omega_{xy}(k_x,k_y)P^{-1} = \Omega_{xy}(-k_x,-k_y)$, $C_{2z}\Omega_{xy}(k_x,k_y)C_{2z}^{-1}=\Omega_{xy}(-k_x,-k_y)$, and $C_{4z}T\Omega_{xy}(k_x,k_y)C_{4z}T^{-1}=-\Omega_{xy}(k_y,-k_x)$. It is therefore clear that the combined operation $C_{4z}T$ simultaneously flips both the valley index and the sign of the Berry curvature, whereas $P$ and $C_{2z}$ leave the Berry curvature invariant. Consequently, systems described by MPG 11.3.37 support the altermagnetic QSHE and exhibit spin-valley locking (SVL), with valley extrema located at the $X$ and $Y$ high-symmetry points.

(ii) Example II: MPG 6.1.17 ($222.1$). The symmetry generators of MPG 6.1.17 are $\{E,\, C_{2x},\, C_{2z}\}$. The rotational symmetries $C_{2x}$ and $C_{2y}$ independently protect spin degeneracy along the high-symmetry lines $\Gamma$–$X$ and $\Gamma$–$Y$, respectively. Assume a valley state $L(\boldsymbol{k}_x,\boldsymbol{k}_y)$ exists along the $\Gamma$–$S$ high-symmetry path. The $C_{2x}$ symmetry operation will generate a symmetry-related counterpart valley $G(k_x,-k_y)$ on the $\Gamma$–$S'$ path, governed by the momentum transformations $C_{2x}L(k_x,k_y)=G(k_x,-k_y)$ and $C_{2z}L(k_x,k_y)=L(-k_x,-k_y)$.
For the Berry curvature, the transformation laws are 
$C_{2x}\Omega_{xy}(k_x,k_y)C_{2x}^{-1}=-\Omega_{xy}(k_x,-k_y)$, and $C_{2z}\Omega_{xy}(k_x,k_y)C_{2z}^{-1}=\Omega_{xy}(-k_x,-k_y)$. This implies that $C_{2x}$ flips both the valley index and the sign of the Berry curvature, while $C_{2z}$ preserves the Berry curvature unchanged.
Additionally, $C_{2x}$ swaps the two constituent atomic layers of the heterostructure, as the $C_{2x}$ rotation can be decomposed into the product of mirror operations $M_y M_z$. Taken together, systems with MPG 6.1.17 can host the altermagnetic QSHE and feature spin-valley-layer locking (SVLL), where the paired valleys are situated at the $L$ and $G$ momentum points.
\begin{table}[htbp]
\caption{List of MPGs and the MLGs that permit altermagnetic QSHE. The symmetry operators critical for realizing the altermagnetic QSHE are listed in the second column denoted by $\mathcal{O}$.}
\label{tab:S1_reordered}
\begin{ruledtabular}
\begin{tabular}{l l p{8.5cm}}
MPG & $\mathcal{O}$ & MLG \\
\hline
$2.1$ & $C_{2\parallel}$ & 8.1.34, 9.1.41, 10.1.45 \\
$m.1$ & $m_{\parallel}$ & 11.1.50, 12.1.57, 13.1.61 \\
$2/m.1$ & $C_{2\parallel}, m_{\parallel}$ & 14.1.66, 15.1.76, 16.1.83, 17.1.90, 18.1.95 \\
$222.1$ & $C_{2\parallel}$ & 19.1.104, 20.1.111, 21.1.118, 22.1.122 \\
$mm2.1$ & $m_{\parallel}$ & 23.1.129, 24.1.136, 25.1.143, 26.1.147, 27.1.154, 28.1.167, 29.1.174, 30.1.181, 31.1.188, 32.1.197, 33.1.202, 34.1.207, 35.1.212, 36.1.221 \\
$mmm.1$ & $C_{2\parallel}, m_{\parallel}$ & 37.1.230, 38.1.243, 39.1.256, 40.1.263, 41.1.276, 42.1.289, 43.1.298, 44.1.307, 45.1.314, 46.1.323, 47.1.330, 48.1.343 \\
$4'$ & $C_{4z}\mathcal{T}$ & 49.3.356 \\
$\bar{4}'$ & $S_{4z}\mathcal{T}$ & 50.3.360 \\
$4'/m$ & $S_{4z}\mathcal{T},C_{4z}\mathcal{T}$ & 51.5.366, 52.5.373 \\
$4'2'2$ & $C_{4z}\mathcal{T},C_{2\parallel}$ & 53.5.378, 54.5.385 \\
$4'm'm$ & $C_{4z}\mathcal{T},m_{\parallel}$ & 55.5.390, 56.5.397 \\
$\bar{4}'2'm$ & $S_{4z}\mathcal{T},m_{\parallel}$ & 57.5.402, 58.5.409 \\
$\bar{4}'m'2$ & $S_{4z}\mathcal{T},C_{2\parallel}$ & 59.5.414, 60.5.421 \\
$4'/mm'm$ & $C_{4z}\mathcal{T},C_{2\parallel}, S_{4z}\mathcal{T},m_{\parallel}$ & 61.9.430, 62.9.443, 63.9.452, 64.9.461 \\
$32.1$ & $C_{2\parallel}$ & 67.1.467, 68.1.470 \\
$3m.1$ & $C_{2\parallel},m_{\parallel}$ & 69.1.473, 70.1.476 \\
$\bar{3}m.1$ & $C_{2\parallel},m_{\parallel}$ & 71.1.479, 72.1.484 \\
$622.1$ & $C_{2\parallel}$ & 76.1.500 \\
$6mm.1$ & $m_{\parallel}$ & 77.1.505 \\
$\bar{6}m2.1$ & $C_{2\parallel}, m_{\parallel}$ & 78.1.510, 79.1.515 \\
$6/mmm.1$ & $C_{2\parallel}, m_{\parallel}$ & 80.1.520 \\
\end{tabular}
\end{ruledtabular}
\end{table}

\section{Effective tight-binding Hamiltonian}
We adopt the tight-binding(TB) Hamiltonian and corresponding hopping parameters reported in Ref.\cite{Feng2025_arXiv} to simulate the altermagnetic QSHE of monolayer $\text{Nb}_2\text{SeTeO}$ [see Fig. S1], which are written as:

\begin{equation}
\begin{aligned}
\mathcal{H}(k_x,k_y) &= \left[\mu + A\left(\cos k_x + \cos k_y\right)\right]\tau_0\sigma_0 
+ B\left[\cos k_x - \cos k_y\right]\tau_z\sigma_0 
+ t\cos\frac{k_x}{2}\cos\frac{k_y}{2}\tau_x\sigma_0 \\
&\quad + \lambda\sin\frac{k_x}{2}\sin\frac{k_y}{2}\tau_y\sigma_z 
+ C\left[\cos k_x - \cos k_y\right]\tau_0\sigma_z 
+ \left[u + D\left(\cos k_x + \cos k_y\right)\right]\tau_z\sigma_z.
\end{aligned}
\label{eq:1}
\end{equation}

Here, $\sigma_i$ and $\tau_i$ denote Pauli matrices corresponding to the spin and sublattice degrees of freedom, respectively. $t$, $t_1$, and $t_2$ stand for nearest-neighbor hopping amplitudes; $\lambda$, $\lambda_1$, and $\lambda_2$ describe spin–orbit coupling (SOC) terms; and $u$ accounts for the on-site local magnetic moment. For simplicity, we introduce two composite hopping parameters $A = t_1 + t_2$ and $B = t_1 - t_2$, which parameterize the isotropic and anisotropic next-nearest-neighbor (NNN) hoppings induced by the effective crystal field from nonmagnetic atomic sites. Analogously, we define $C = \lambda_1 + \lambda_2$ and $D = \lambda_1 - \lambda_2$ to characterize the NNN off-site SOC contributions.

\begin{figure}[htbp]
\centering
\includegraphics[width=4in]{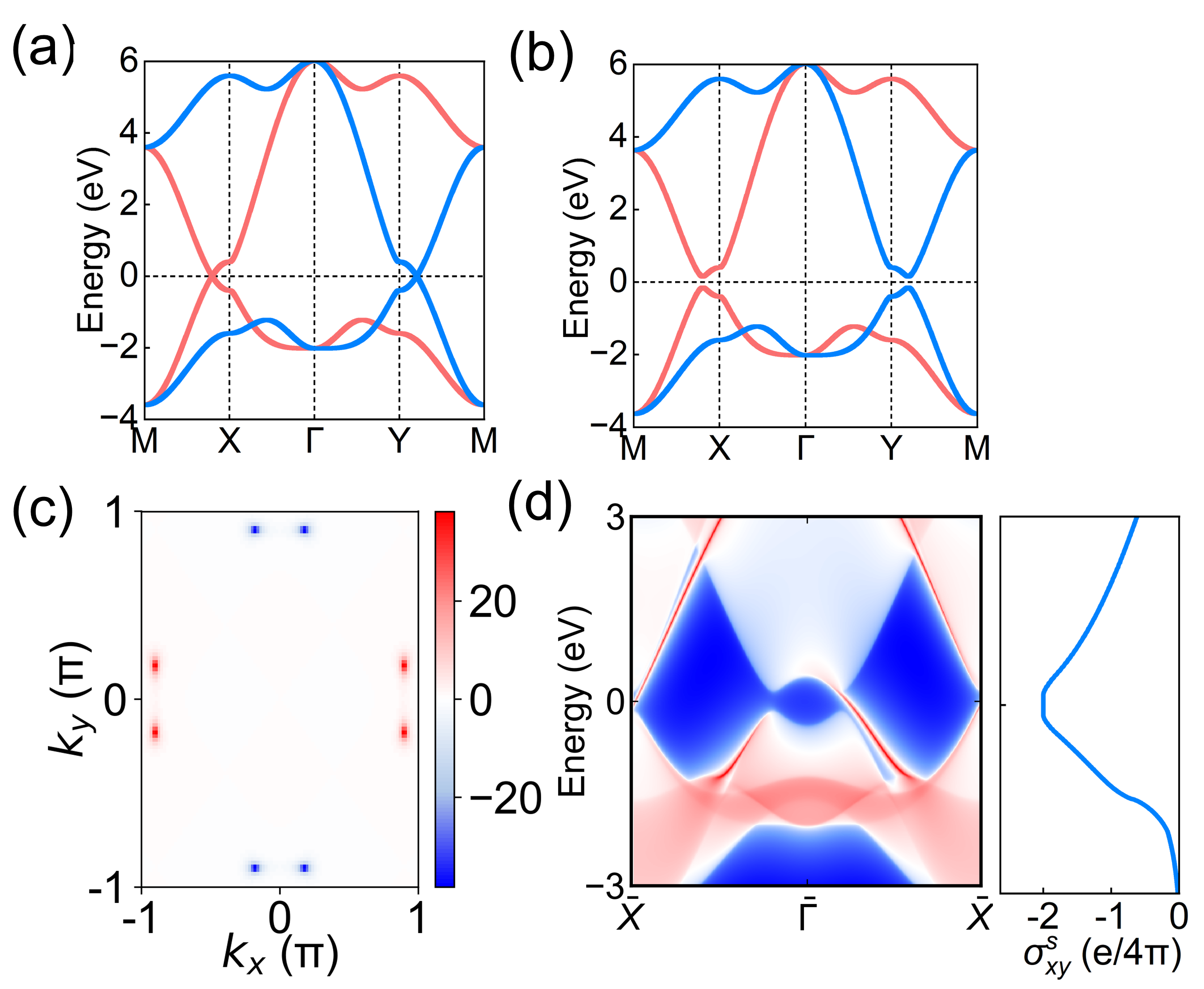} 
\vspace{-1em}%
\caption{Band structures for (a) $\lambda = 0$, and (b) $\lambda = 0.5$. The remaining parameters are fixed to $A = 0.5$, $B = -1$, $C = 0.5$, $D = 1$, $\lambda = 0.5$, $u = -1.6$, $\mu = 1$, and $t = 4$. (c) Berry curvature for (b), (d) Edge states and SHC for (b).}
\label{fig:label1}
\end{figure}

\begin{figure}[htbp]
\centering
\includegraphics[width=4in]{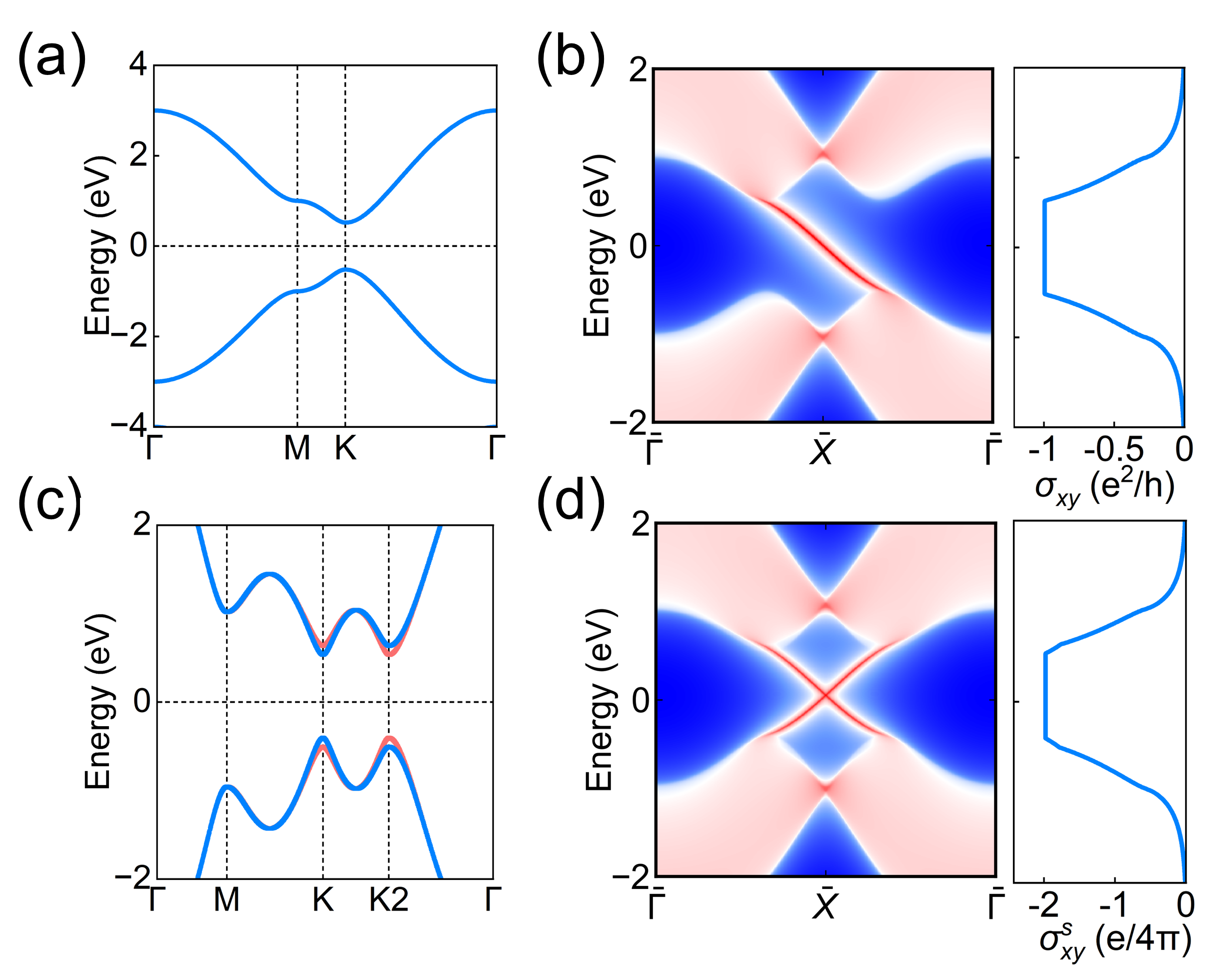} 
\vspace{-1em}%
\caption{(a) Band structures, (b) edge states and anomalous Hall conductivity (AHC) for monolayer Hamiltonian.  (c) Band structures, (d) edge states and SHC for bilayer Hamiltonian. The parameters are fixed to $t_{\text{on}} = -7t,\ \lambda = 3.5t,\ t_{\text{SO}} = 0.1t,\ t_1 = 0.3t,\ t_2 = 0.1t,\ t_3 = 0.1t$.}
\label{fig:label2}
\end{figure}

For the altermagnetic QSHE calculations of  $\text{Hf}_3\text{Se}_3\text{Te}_2$, we utilize the TB model and its full set of hopping parameters from Ref.\cite{Niu2025l} [see Fig. S2], expressed as follows:
\begin{equation}
\begin{aligned}
H_{\text{monolayer}} &= t_{\text{on}} \sum_{i} c_i^\dagger c_i 
+ t \sum_{\langle ij \rangle} c_i^\dagger c_j 
+ i t_{\text{SO}} \sum_{\langle\langle ij \rangle\rangle} v_{ij} c_i^\dagger s_z c_j \\
&\quad + \lambda \sum_{i} c_i^\dagger B_z s_z c_i.
\end{aligned}
\label{eq:2}
\end{equation}

\begin{equation}
\begin{aligned}
H_{\text{bilayer}} &= t_{\text{on}} \sum_{i} c_i^\dagger c_i 
+ t \sum_{\langle ij \rangle} c_i^\dagger c_j 
+ i t_{\text{SO}} \sum_{\langle\langle ij \rangle\rangle} v_{ij} c_i^\dagger s_z c_j \\
&\quad + \lambda \sum_{i} c_i^\dagger B_z s_z c_i 
+ \sum_{\langle\langle ij \rangle\rangle} \big(t_1 c_i^\dagger c_j + t_2 c_i^\dagger c_j\big) \\
&\quad + t_3 \sum_{i} c_{i,a}^\dagger c_{i,b}.
\end{aligned}
\label{eq:3}
\end{equation}

Here, the parameters representing on-site energy, nearest-neighbor (NN) hopping strength, intrinsic SOC, and ferromagnetic exchange field are denoted by $t_{\text{on}}$, $t$, $t_{\text{SO}}$, and $\lambda$, respectively. $\nu_{ij} = \pm 1$ corresponds to counterclockwise and clockwise hopping paths, and $s_z$ stands for the $z$-component Pauli spin matrix. For the bilayer system, we further introduce the interlayer coupling $t_3$ between the top and bottom atomic layers, together with two distinct next-nearest-neighbor (NNN) hopping amplitudes: $t_1$ for intrasublattice hoppings within sublattices A/C and $t_2$ for intrasublattice hoppings within sublattices B/D.

\section{Brillouin Zones for models and materials in the Main Text}
Figure S3 shows the Brillouin zones (BZs) of the models and materials discussed in main text. Specifically, the
BZ of the lattice model (5) in the main text is presented in Fig. S3(a), The BZs of monolayer $\text{Nb}_2\text{SeTeO}$ and monolayer $\text{Hf}_3\text{Se}_3\text{Te}_2$ are shown as Fig.
S3(b) and Fig. S3(c), respectively.

\begin{figure}[htbp]
\centering
\includegraphics[width=6in]{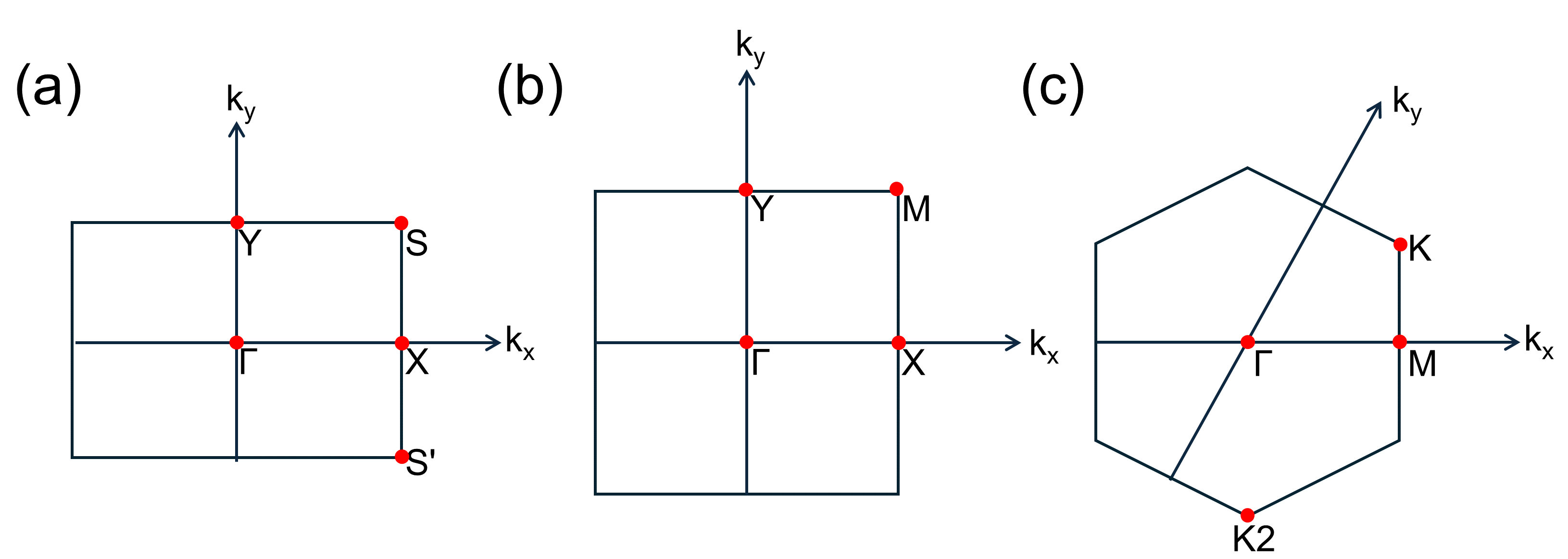} 
\vspace{-1em}%
\caption{(a) BZ of lattice model (5) in the main text. BZs of (b) monolayer $\text{Nb}_2\text{SeTeO}$ and (c) monolayer $\text{Hf}_3\text{Se}_3\text{Te}_2$.}
\label{fig:label3}
\end{figure}

\section{Computational methond}
All first-principles calculations were performed using the projector augmented wave (PAW)~\cite{Kresse1999} method based on density functional theory (DFT) as implemented in the Vienna Ab initio Simulation Package (VASP)~\cite{Kresse1993,Kresse1996}. The exchange-correlation potential was treated using the generalized gradient approximation (GGA) with the Perdew-Burke-Ernzerhof (PBE) functional~\cite{Hohenberg1964,Kohn1965,Ceperley1980,Perdew1981,Perdew1996}. A plane-wave cutoff energy was set to 550 eV. The convergence criteria for energy and force were set to $10^{-6}$ eV and $0.01$ eV/\AA, respectively. A vacuum slab of 20 \AA\ was applied to prevent spurious interactions between periodic images. For geometry optimization, the first Brillouin zone was sampled by a $\Gamma$-centered $k$-mesh of $13 \times 13 \times 1$~\cite{Monkhorst1976}. To account for the correlation effects of $3d$ electrons, the PBE+$U$ method with $U_{\text{eff}} = 4$ eV was employed~\cite{Anasori2016,Zhang2018,Yao2022}. Phonon dispersions were calculated using $3 \times 3 \times 1$ supercells via the Phonopy code~\cite{Togo2015} based on density functional perturbation theory (DFPT). The topological edge states and spin Hall conductivity (SHC) were calculated using the Wannier90~\cite{Mostofi2008} and WannierTools~\cite{Wu2018} packages.

Berry curvature is derived from the equation:
\begin{equation}\label{eq:4}
\Omega(\mathbf{k}) = -\sum_n \sum_{n\neq m} \frac{f_n(\mathbf{k}) \times 2\mathrm{Im} \langle \psi_{n\mathbf{k}} | \hat{v}_x | \psi_{m\mathbf{k}} \rangle \langle \psi_{m\mathbf{k}} | \hat{v}_y | \psi_{n\mathbf{k}} \rangle}{(E_{n\mathbf{k}} - E_{m\mathbf{k}})^2}
\end{equation}
where $\mathbf{k}$ is the electron wave vector and $f_n(\mathbf{k})$ defines as the Fermi-Dirac distribution function, $\hat{v}_x$ and $\hat{v}_y$ are velocity operators of the Dirac electrons, $n$ and $m$ are the band indexes, $E_{n\mathbf{k}}$ and $E_{m\mathbf{k}}$ are the eigenvalues of the Bloch wave functions $\psi_{n\mathbf{k}}$ and $\psi_{m\mathbf{k}}$, respectively.

The AHC could be calculated from below:
\begin{equation}\label{eq:5}
\sigma_{xy} = -\frac{e^2}{\hbar} \int_{\text{BZ}} \frac{d^3\mathbf{k}}{(2\pi)^3} \Omega(\mathbf{k})
\end{equation}
where BZ defines as the first Brillouin zone, $\mathbf{k}$ is the electron wave vector, $\Omega(\mathbf{k})$ is Berry curvature.

The SHC is calculated by the following formula:
\begin{equation}\label{eq:6}
\sigma_{xy}^{s_z}(\omega) = \hbar \int_{\text{BZ}} \frac{d^3\mathbf{k}}{(2\pi)^3} \sum_n f_{n\mathbf{k}} \times \sum_{m\neq n} \frac{2\mathrm{Im}\left[\langle n\mathbf{k}|\hat{j}_{x}^{s_z}|m\mathbf{k}\rangle \langle m\mathbf{k}|-e\hat{v}_y|n\mathbf{k}\rangle\right]}{(\varepsilon_{n\mathbf{k}} - \varepsilon_{m\mathbf{k}})^2 - (\hbar\omega + i\eta)^2}
\end{equation}
where $n$ and $m$ are the band indexes, $\varepsilon_n$ and $\varepsilon_m$ are the eigenvalues, $\mathbf{k}$ is the electron wave vector, $f_{n\mathbf{k}}$ is the Fermi distribution function, $\hat{j}_{x}^{s_z}$ is the spin current operator in the projection of spin $z$ direction ($s_z$), $\hat{v}_y$ is the velocity operator and both the angular frequency $\omega$ and the damping factor $\eta$ are reduced to zero in the scenario of a direct current with a clean limit.

\section{Other Candidate Materials}

Monolayer $\mathrm{Ti}_2\mathrm{Te}_2\mathrm{O}$ is a square structure with layer group of P4/mmm (No. 61) [see Fig. S4(a)]. After optimization, the lattice constants of monolayer $\mathrm{Ti}_2\mathrm{Te}_2\mathrm{O}$ are a = b = 4.25 Å. In the absence of SOC, monolayer $\mathrm{Ti}_2\mathrm{Te}_2\mathrm{O}$ exhibits non-relativistic spin splitting, with one pair of Weyl points of opposite spin emerging located ${\Gamma}$ points [see Fig. S4(b)]. When SOC is considered, band inversion occurs at the Weyl points[see Fig. S4(c)], and belongs to MPG 15.4.56. The edge state of monolayer $\mathrm{Ti}_2\mathrm{Te}_2\mathrm{O}$ reveals two topologically protected edge states with opposite chirality near the $\tilde{\Gamma}$ points, as shown in Fig. S4(d). Meanwhile, our calculations demonstrate that the SHC of monolayer $\mathrm{Ti}_2\mathrm{Te}_2\mathrm{O}$ is quantized as $|\sigma_{xy}^{s}|=2e/4\pi$. 

\begin{figure}[htbp]
\centering
\includegraphics[width=4in]{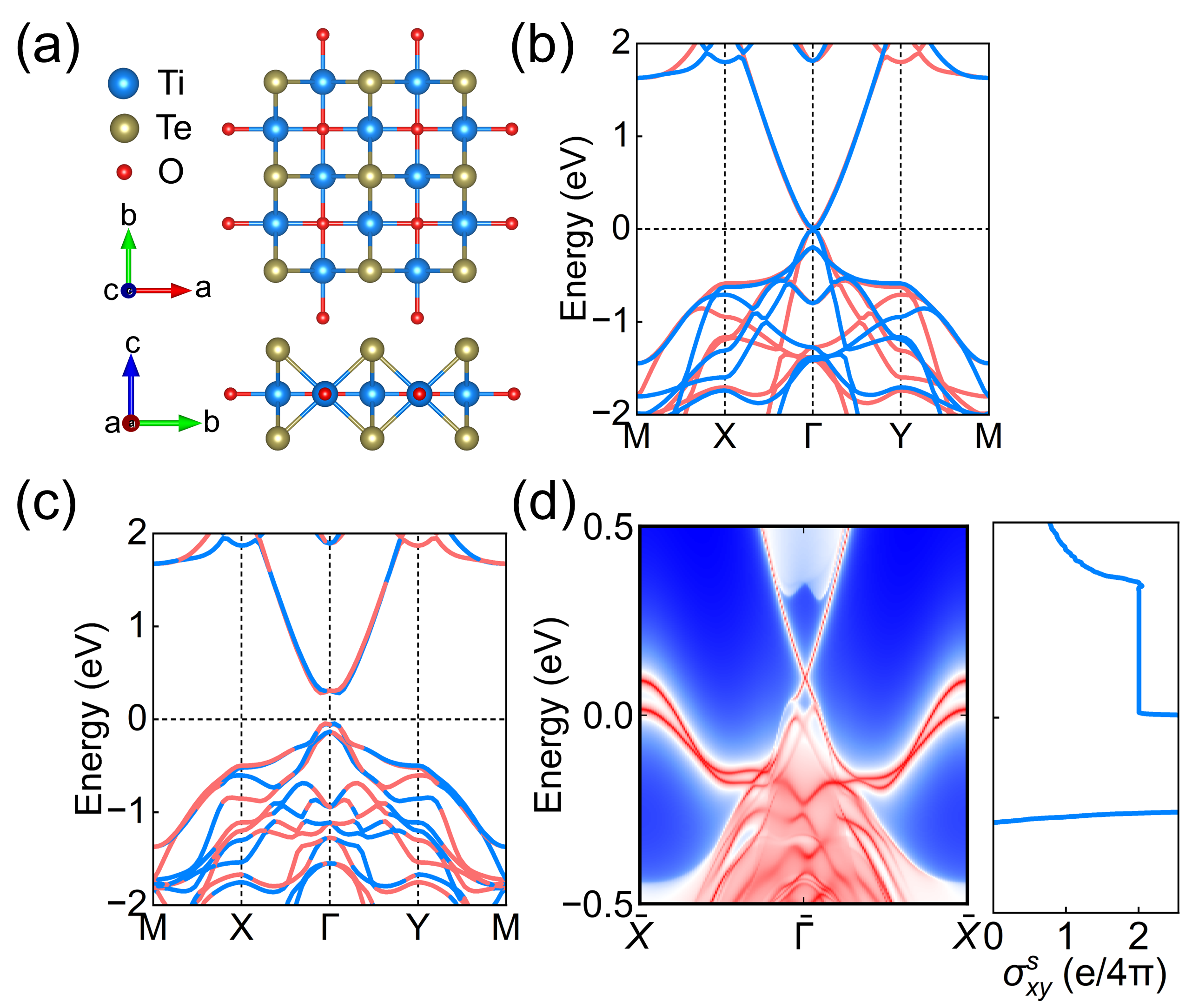} 
\vspace{-1em}%
\caption{(a) Top view and side view of monolayer $\mathrm{Ti}_2\mathrm{Te}_2\mathrm{O}$.
(b) The band structure without SOC.
(c) The band structure with SOC. 
(d) The edge states and SHC of monolayer $\mathrm{Ti}_2\mathrm{Te}_2\mathrm{O}$.}
\label{fig:label4}
\end{figure}

\section{supplemental figures}

In order to confirm the magnetic ground state, four magnetic patterns of the $2\times2\times1$ supercell of $\text{Nb}_2\text{SeTeO}$ are constructed in Fig.~S5. The total energy calculations of these four magnetic configurations demonstrate that the AM state shown in Fig.~S5(b) possesses the lowest energy and thus is the magnetic ground state. The calculated total energies are listed as follows: FM = 1.16 eV, AM = 0 eV, AFM1 = 1.30 eV, AFM2 = 0.82 eV. In addition, the monolayer $\mathrm{Hf}_3\mathrm{Se}_3\mathrm{Te}_2$ hosts a FM ground state [see Fig.~S6]. The calculated total energies are listed as follows: FM = 0 meV,  AFM1 = 156 meV, AFM2 = 59 meV.

\begin{figure}[htbp]
\centering
\includegraphics[width=4in]{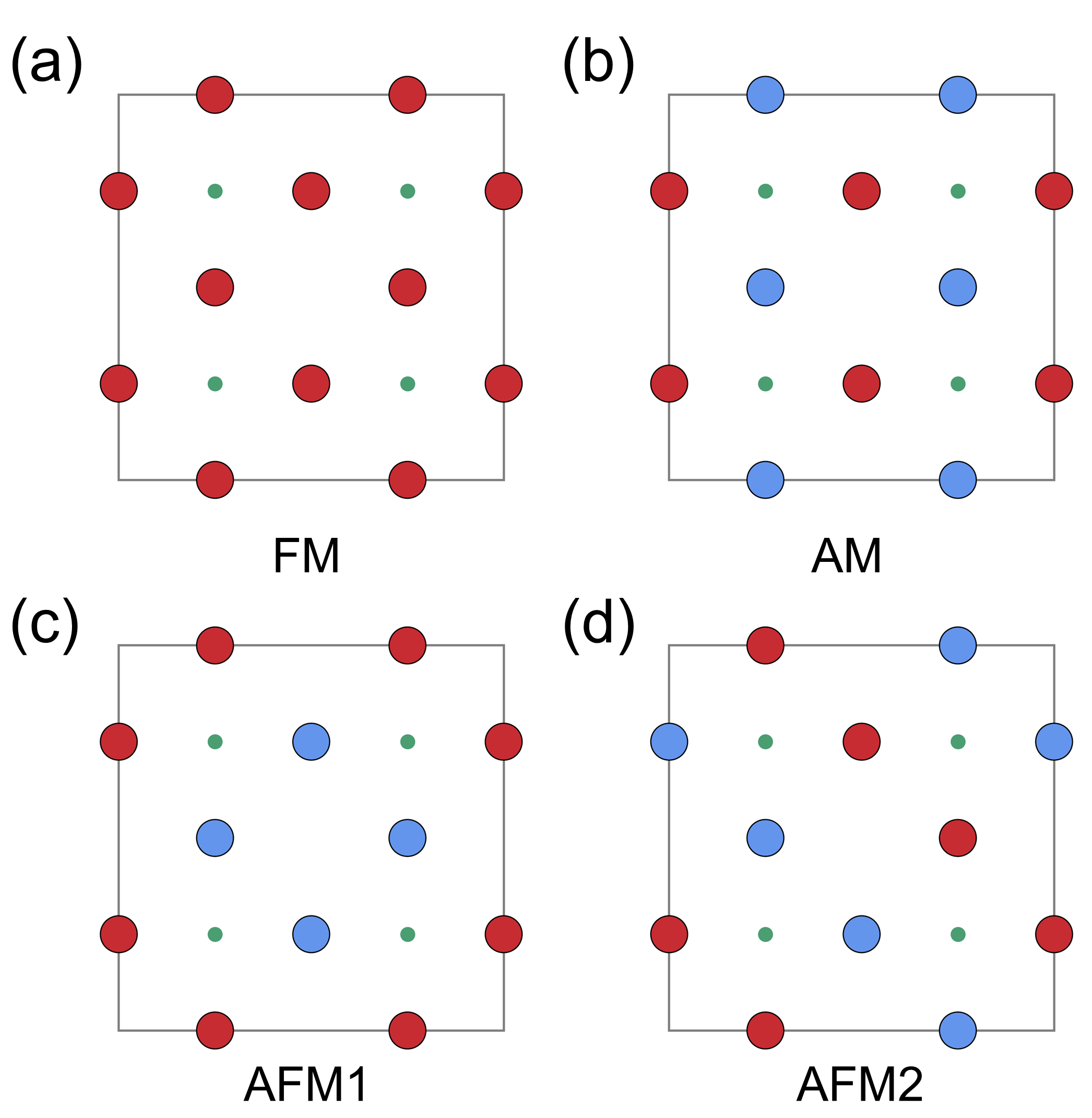} 
\vspace{-1em}%
\caption{Distinct magnetic configurations  for (a) FM,
(b) AM, 
(c) AFM1, (d) AFM2 of monolayer $\mathrm{Nb}_2\mathrm{Se}\mathrm{Te}\mathrm{O}$ and monolayer $\mathrm{Ti}_2\mathrm{Te}_2\mathrm{O}$.}
\label{fig:labe5}
\end{figure}

\begin{figure}[htbp]
\centering
\includegraphics[width=6in]{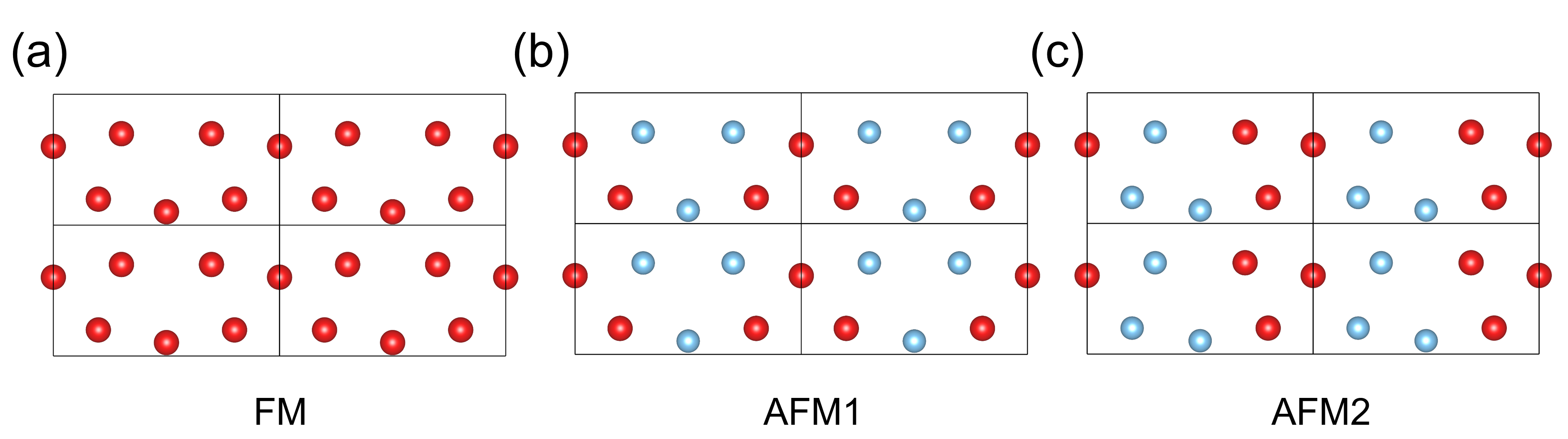} 
\vspace{-1em}%
\caption{Distinct magnetic configurations for (a) FM, 
(b) AFM1, (c) AFM2 of monlayer $\mathrm{Hf}_3\mathrm{Se}_3\mathrm{Te}_2$.}
\label{fig:labe6}
\end{figure}

\begin{figure}[htbp]
\centering
\includegraphics[width=6in]{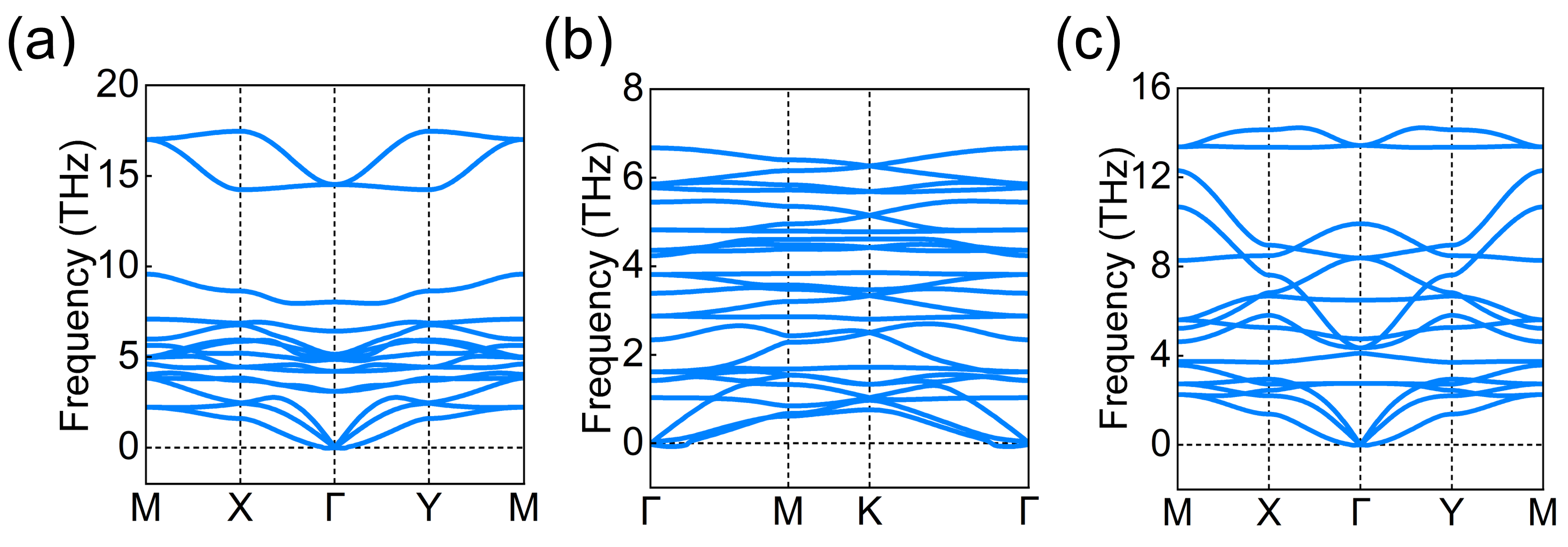} 
\vspace{-1em}%
\caption{The calculated phonon spectrum of (a) monolayer $\mathrm{Nb}_2\mathrm{Se}\mathrm{Te}\mathrm{O}$, 
(b) monolayer $\mathrm{Hf}_3\mathrm{Se}_3\mathrm{Te}_2$, (c) monolayer $\mathrm{Ti}_2\mathrm{Te}_2\mathrm{O}$.}
\label{fig:labe7}
\end{figure}

\begin{figure}[htbp]
\centering
\includegraphics[width=4in]{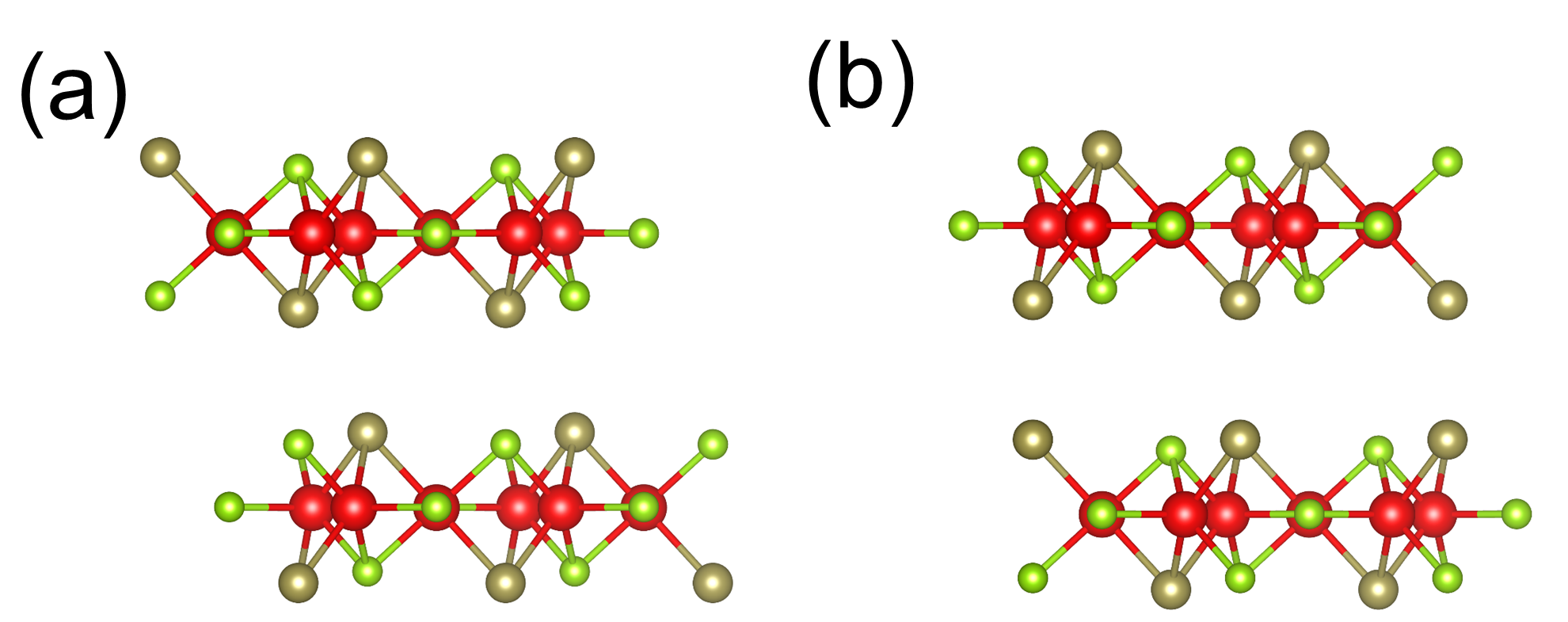} 
\vspace{-1em}%
\caption{side view for (a) AA stacked $\mathrm{Hf}_3\mathrm{Se}_3\mathrm{Te}_2$, 
(b) AB stacked $\mathrm{Hf}_3\mathrm{Se}_3\mathrm{Te}_2$.}
\label{fig:labe8}
\end{figure}

\begin{figure}[htbp]
\centering
\includegraphics[width=4in]{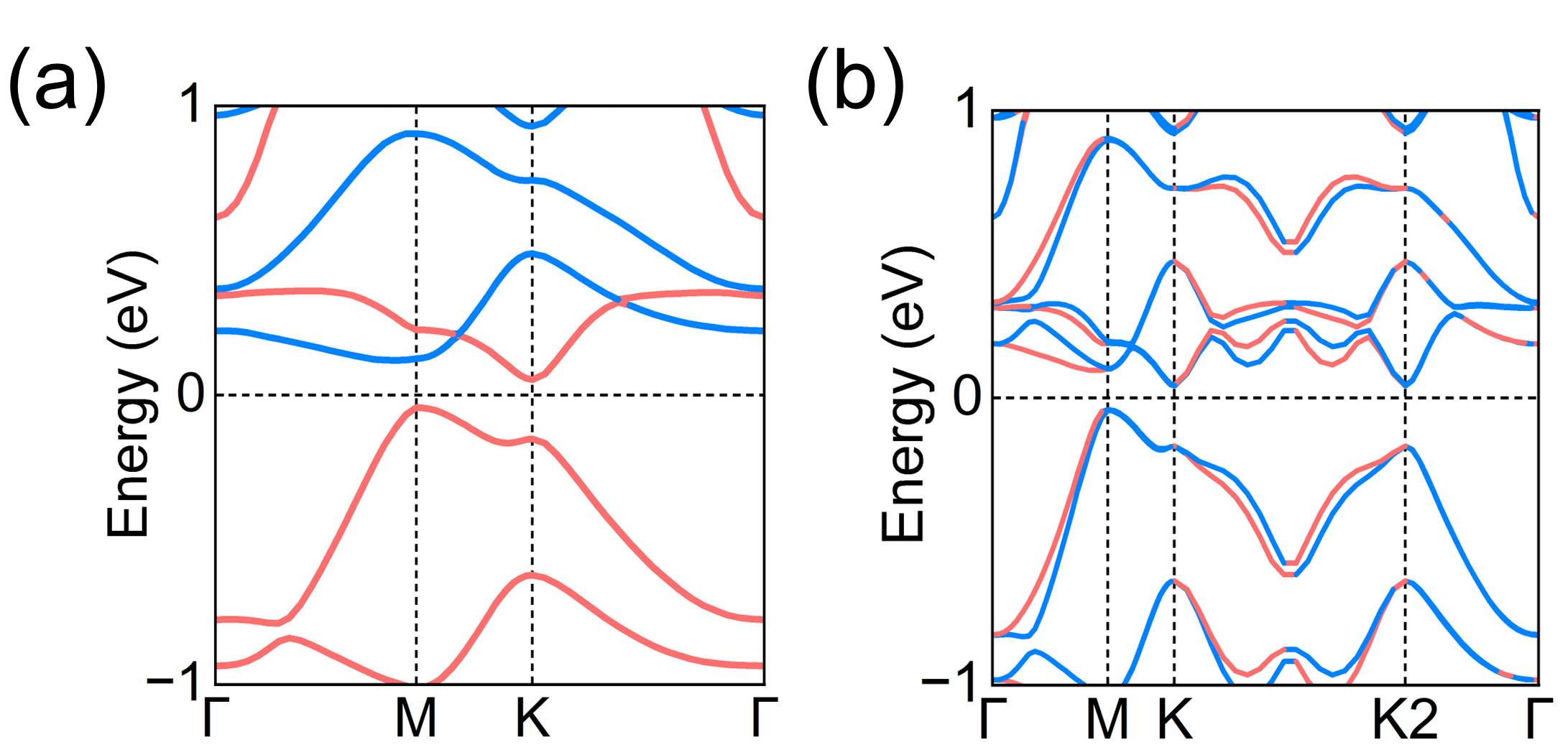} 
\vspace{-1em}%
\caption{(a) Band structure with SOC for monolayer $\mathrm{Hf}_3\mathrm{Se}_3\mathrm{Te}_2$, 
(b) Band structure with SOC for AB stacked $\mathrm{Hf}_3\mathrm{Se}_3\mathrm{Te}_2$.}
\label{fig:labe9}
\end{figure}

\begin{figure}[htbp]
\centering
\includegraphics[width=4in]{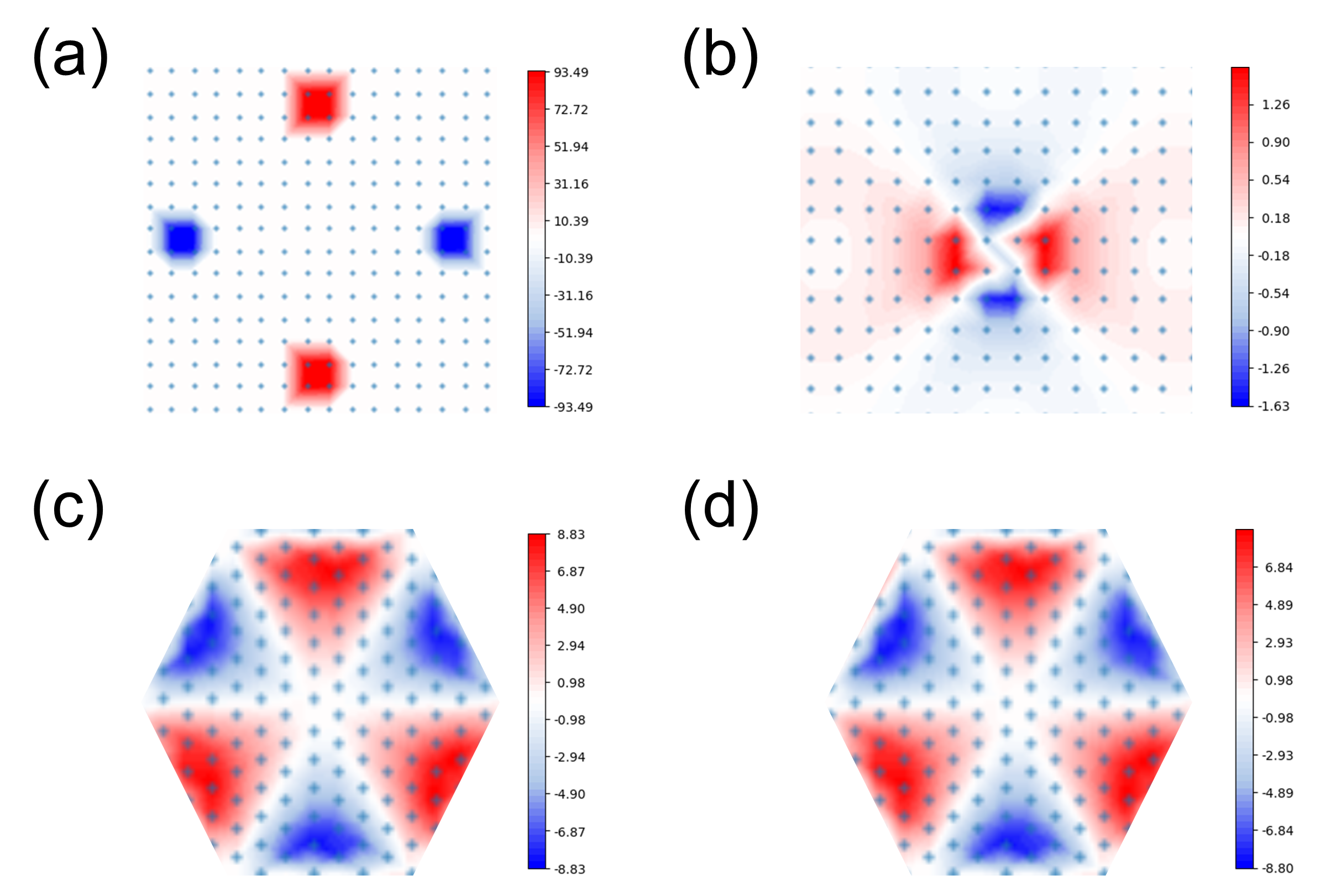} 
\vspace{-1em}%
\caption{Berry curvature for (a) monolayer $\mathrm{Nb}_2\mathrm{Se}\mathrm{Te}\mathrm{O}$, 
(b) monolayer $\mathrm{Ti}_2\mathrm{Te}_2\mathrm{O}$, (c) AA stacked $\mathrm{Hf}_3\mathrm{Se}_3\mathrm{Te}_2$,
(d) AB stacked $\mathrm{Hf}_3\mathrm{Se}_3\mathrm{Te}_2$.}
\label{fig:labe10}
\end{figure}

\begin{figure}[htbp]
\centering
\includegraphics[width=4in]{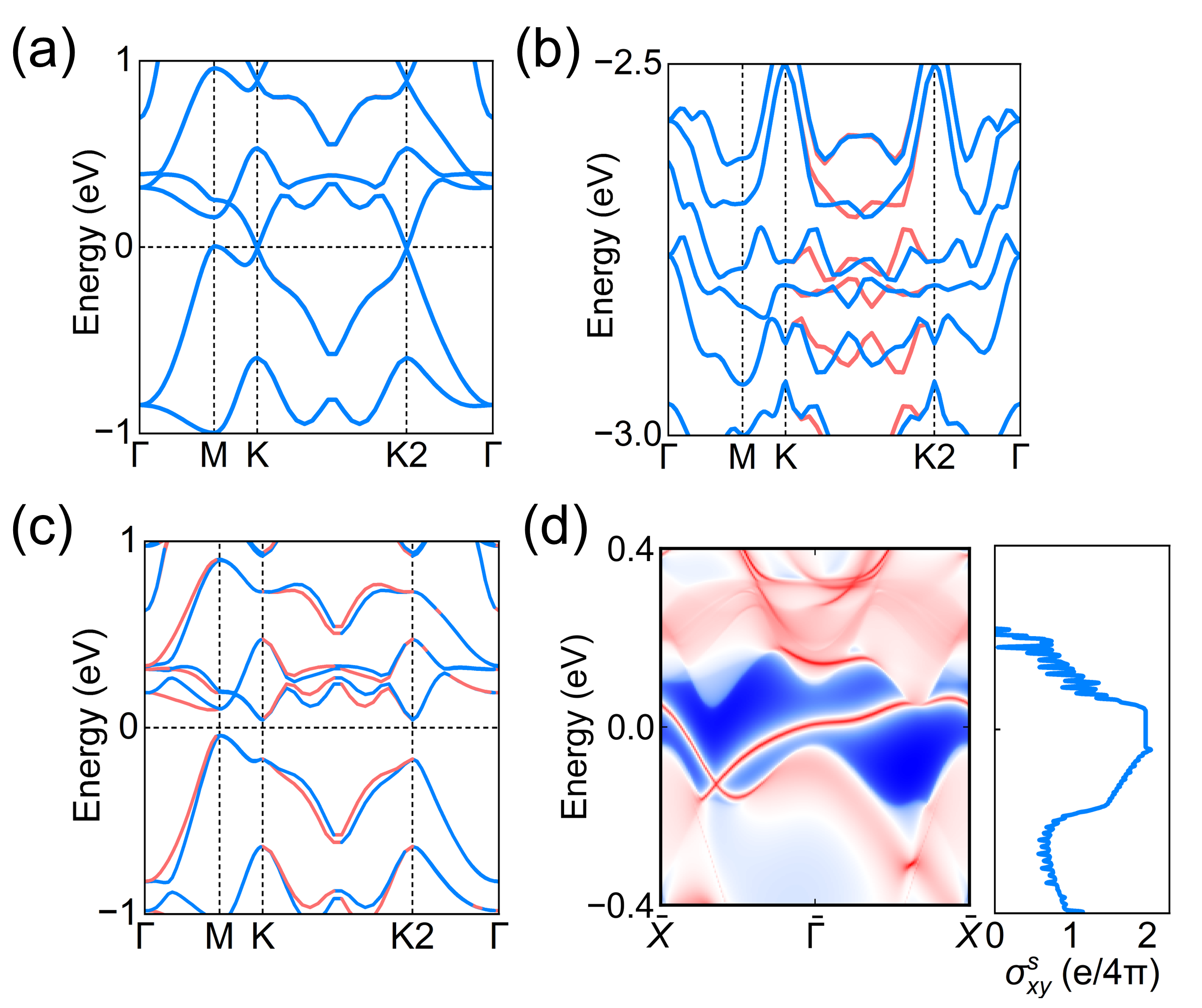} 
\vspace{-1em}%
\caption{(a) Band structure without SOC for AB stacked $\text{Hf}_3\text{Se}_3\text{Te}_2$.
(b) Zoom-in band structure  without SOC for AB stacked $\text{Hf}_3\text{Se}_3\text{Te}_2$.
(c) Band structure with SOC for AB stacked $\text{Hf}_3\text{Se}_3\text{Te}_2$.
(d) The edge states and SHC for AB stacked  $\text{Hf}_3\text{Se}_3\text{Te}_2$ along the [100] direction.}
\label{fig:labe11}
\end{figure}

\bibliography{supref}